\documentclass{aa}
\usepackage{psfig}
\usepackage{amssymb}

\newcommand{\teff}{\ifmmode T_{\rm eff} \else $T_{\mathrm{eff}}$\fi}
\newcommand{\logg}{\ifmmode \log g \else $\log g$\fi}
\newcommand{\msun}{\ifmmode M_{\odot} \else M$_{\odot}$\fi}
\newcommand{\zsun}{\ifmmode Z_{\odot} \else Z$_{\odot}$\fi}
\newcommand{\lsun}{\ifmmode L_{\odot} \else L$_{\odot}$\fi}

\newcommand{\mdot}{\dot{M}}
\newcommand{\myr}{M$_{\odot}$ yr$^{-1}$}
\newcommand{\vsini}{$V$ sin$i$}

\newcommand{\civ}{C~{\sc iv} $\lambda\lambda$1548,1551 }
\newcommand{\nv}{N~{\sc v}  $\lambda$1238,1242 }
\newcommand{\niv}{N~{\sc iv}  $\lambda$1718 }
\newcommand{\oiv}{O~{\sc iv} $\lambda\lambda$1339,1343 }
\newcommand{\ov}{O~{\sc v}  $\lambda$1371 }
\newcommand{\ciii}{C~{\sc iii} $\lambda \lambda 1426,1428$ }

\def\aap{A\&A}

\def\aj{AJ}
\def\apj{ApJ}

\def\apjs{ApJS}
\def\mnras{MNRAS}

\begin{document}


\title{Puzzling wind properties of young massive stars in SMC-N81}

\author{Fabrice Martins 
           \inst{1,2}
        \and Daniel Schaerer 
           \inst{2,1}
        \and D. John Hillier
           \inst{3}
        \and Mohammad Heydari-Malayeri 
	   \inst{4}}

\offprints{F. Martins, fabrice.martins@obs.unige.ch \inst{1}}

 \institute{Laboratoire d'Astrophysique, Observatoire Midi-Pyr\'en\'ees,
 14 Av.  E. Belin, F-31400 Toulouse, France 
 \and
 Observatoire de Gen\`eve, 51 Chemin des Maillettes, CH-1290 Sauverny, 
 Switzerland
 \and 
 Department of Physics and Astronomy, University of Pittsburgh, 3941
 O'Hara Street, Pittsburgh, PA 15260, USA
 \and
 LERMA, Observatoire de Paris, 61 Avenue de l'Observatoire, F-75012
 Paris, France
}

\date{Recieved 14 October 2003 / Accepted 16 March 2004}

\titlerunning{Young O stars in SMC N81}
\titlerunning{Young massive stars in SMC N81}


\abstract{
We present a quantitative study of massive stars in the High
Excitation Blob N81, a compact star forming region in the SMC.
The stellar content
was resolved by HST and STIS was used to obtain medium
resolution spectra.
The qualitative analysis of the stellar properties presented in
Heydari-Malayeri et al.\ (\cite{papI}) is 
extended using 
non-LTE spherically extended atmosphere models including
line-blanketing computed with the code CMFGEN (Hillier \& Miller
\cite{hm98}), and the wind properties are investigated. The main results
are the following: 

\begin{itemize}
\item The SMC-N81 components are young ($\sim$ 0--4
Myrs) O stars with effective temperatures compatible with
medium to late 
subtypes and with luminosities lower than average 
Galactic O dwarfs, rendering them possible ZAMS candidates. 

\item The winds are extremely weak: with values of the order
 10$^{-8}$/10$^{-9}$ \myr\, the mass loss rates are lower than observed so 
 far for Galactic dwarfs. Only the recent study of SMC stars by Bouret
 et al.\ (\cite{jc03}) show the same trend. The modified wind momenta
 ($\dot{M}$ v$_{\infty}$ $\sqrt{R}$) are also 1 to 2 orders of
 magnitude lower than observed for Galactic stars. Both the mass loss
 rates and the modified wind momenta are lower than the predictions of 
 the most recent hydrodynamical models.
\end{itemize}

The accuracy of the UV based mass loss rate determination, relying  
in particular on the predicted ionisation fractions, are carefully
examined. We find that $\dot{M}$ could be underestimated by a factor
of up to 10. Even in this unlikely case, the above conclusions
remain valid  \textit{qualitatively}. 

The reasons for such weak winds are
investigated with special emphasis on the modified wind momenta:

\begin{itemize}
\item There may be a break-down of the wind momentum - luminosity
relation (WLR) for dwarf stars at low luminosity (log L/L$_{\odot}$
$\la$ 5.5). However, reasons for such a
breakdown remain unknown. 

\item The slope of the WLR may be steeper at low metallicity. This is
predicted by the radiation driven wind theory, but the current
hydrodynamical simulations do not show any change of the slope at SMC
metallicity. Moreover, there are indications that some Galactic objects
have wind momenta similar to those of the SMC stars.

\item Decoupling may take place in the atmosphere of the SMC-N81
stars, leading to multicomponent winds. However, various tests
indicate that this is not likely to be the case.
\end{itemize}

  The origin of the weakness of the wind observed in the SMC-N81
  stars remains unknown. 
  We suggest that this weakness may be linked with the youth of 
  these stars and represents possibly the onset of stellar winds in 
  recently formed massive stars.

}

\maketitle
\section{Introduction}
\label{s_intro}
Massive stars play key roles in various astrophysical contexts all
along their evolution: they ionise ultra-compact HII regions while
still embedded in their parental molecular cloud; they create ionised
cavities and shape the surrounding interstellar medium
during the main fraction of their lifetime;
they experience
strong episodes of mass loss when they become Luminous Blue Variables
and Wolf-Rayet stars, revealing their core and enriching the ISM in
products of H and He burning; they end their life as
supernovae, producing the heavy elements and releasing large
amounts of mechanical energy. 
During all these phases, massive stars lose mass through winds driven
by radiation pressure on metallic lines. This affects not only
their evolution (e.g. Chiosi \& Maeder \cite{cm86}) but also the
surrounding interstellar medium in which the release of mechanical
energy can trigger instabilities leading to the collapse of molecular
clouds and to star formation. Moreover, bubbles and superbubbles
observed on galactic scales are powered by such mass ejections. Hence, 
various astrophysical
fields require the knowledge of quantitative wind properties of
massive stars. 

Several studies have been carried out in the last two decades to
determine these properties.
At solar metallicity, the observational determinations 
(e.g. Howarth \& Prinja \cite{hp89}, Puls et al.\ \cite{puls96},
Herrero et al.\ \cite{hpv00}) are on average in good agreement with
the most recent hydrodynamical predictions based on the
radiation driven wind theory (Vink et al.\ \cite{vink00}), both in
terms of mass loss rate and of the modified wind momentum - 
luminosity relation (WLR,
e.g. Puls et al.\ \cite{puls96}) which quantifies the strength of the wind. 
At non solar metallicities, we expect the wind properties to vary with 
$Z$ due to the modified radiative acceleration through metallic lines. In
particular, the mass loss rate should be proportional to
$Z^{r}$ (Abbott \cite{abbott82}, Puls et al.\ \cite{psl00}) and the WLR 
should be shifted towards lower values and should have a steeper
slope. The most recent theoretical results
predict $r \sim 0.8$ (Vink et al.\ \cite{vink01}) but no change in the 
slope of the WLR, at least for $Z > 10^{-3} Z_{\odot}$ 
(Hoffmann et al.\ \cite{tadziu02}, Kudritzki
\cite{kud02}). Observational studies indicate a reduction of the mass
loss rate and of the terminal velocity in the Magellanic Clouds, but
given the small number of objects studied so far,
the behaviour of the WLR at low metallicity is still poorly understood.
Several groups are currently analysing stars in
sub solar (Crowther et al.\ \cite{paul02}, Hillier
et al.\ \cite{hil03}, Bouret et al.\ \cite{jc03}) and super solar
(Najarro et al.\ \cite{paco} , Figer et al.\ \cite{figer})
regions for a better understanding of wind properties in different
environments. The present work on SMC-N81 stars takes part in this
effort. 

The SMC-N81 region belongs to the class of the ``High Excitation Blobs''
(HEB) first introduced by Heydari-Malayeri \& Testor
(\cite{ht82}). These blobs are compact regions of star formation in
the Magellanic Clouds (see Heydari-Malayeri \cite{mhmiau} for a
complete review). They have a typical radius of a few pc and display the
features of star forming regions: HII cavities, turbulent
structures, ionisation fronts and shocks. 
Recent HST observations
(Heydari-Malayeri et al.\ \cite{pap0}) have revealed for the
first time its stellar content, 4 of the brightest stars being grouped 
in the central 2 \arcsec\ wide region.
Subsequently spectra of the main
exciting stars have been obtained with STIS onboard HST. 
The qualitative analysis of these spectra, presented in
Heydari-Malayeri et al.\ (\cite{papI}, hereafter paper I), have already
revealed interesting
properties. First, the stars have been identified as mid O dwarfs 
with surprisingly low luminosities compared to ``classical''
dwarfs. Second, the UV spectra have shown signatures of very weak
winds, even weaker than those usually observed in the SMC. These
characteristics have lead Heydari-Malayeri et al.\ (\cite{papI}) to
propose that the 
SMC-N81 stars could belong to the class of Vz stars which are massive
stars thought to lie very close to the ZAMS (Walborn \& Parker
\cite{wp92}). 

As such the properties of these stars, showing unusually weak winds
compared to other SMC O stars, seem already quite interesting.
Furthermore the association of these objects with a compact 
star forming region, presumably indicative of a very young age,
allows one also to obtain unique constraints on properties 
of very young massive stars shortly after their birth.
In fact such observations appear
crucial for a better understanding of the earliest evolutionary phases 
of massive stars and to constrain their formation
process which is still under debate (they may form by accretion on a
protostellar core -- Norberg \& Maeder \cite{nm}, Behrend \& Maeder
\cite{raoul}-- or by collisions between low mass components in dense
stellar clusters -- Bonnell et al.\ \cite{bonnell}).

With such objectives in mind we have carried out a quantitative 
study of the UV spectra of the SMC-N81 stars. 
First results have been presented in Martins et al.\ (\cite{lanzarote}).
In fact we are able to determine upper limits on the mass loss
rates of four O stars in this region, which turn out to be
surprisingly low (typically $\mdot \la$ a few $10^{-9}$ \myr) compared
to predictions of the radiation driven wind theory, even
when taking metallicity effects into account.
Although no precise physical explanation is found for this 
behaviour we strongly suggest that this behaviour is related
to the very youth of these massive stars.

The remainder of the paper is structured as follows.
Section \ref{observations} briefly summarises the observations and data
reduction. Section \ref{cmfgen} describes the main ingredients of the
modeling. In Sect.\ \ref{interstellar} we explain how 
interstellar lines are taken into account. The main results
are given in Sect.\ \ref{analysis} and discussed in
Sect. \ref{neb_ste_prop} (nebular and stellar properties) and 
\ref{wind_properties} (wind properties). Finally, Sect.\
\ref{conclusion} summarises the main results.


\section{Observations and data reduction}

\label{observations}
Ten SMC N81 stars were observed 
with STIS onboard HST on 28 and 31 October
1999 (General Observer Program No 8246, PI M.\ Heydari-Malayeri). 
The spectra in the wavelength 
range 1120 - 1715 \AA\ where obtained through the G140 L grating on the 
Multi-Anode Microchannel Array (MAMA) detector. The 52''$\times$0.2''
entrance slit was used. The effective resolution was 0.6 \AA\ per pixel 
of 25 $\rm{\mu}$m which correspond to a dispersion of 24 \AA\ mm$^{-1}$ or
a resolution of 1.2 \AA. The exposure times were chosen to equalize
the S/N ratios which are of the order 20 for the 4 stars studied here.
The other ones have lower S/N ratio which precludes any quantitative
analysis. 
Optical spectra, also obtained with STIS for 6 objects, were of 
insufficient quality (too low spectral resolution) and are therefore
not used here.

Details concerning the data reduction can be found in paper I.


\section{CMFGEN: basic concepts}
\label{cmfgen}
The modeling of realistic atmospheres of massive stars requires the
inclusion of three main ingredients: 1)
due to the high luminosity of these stars, radiative processes are
dominant and a non-LTE treatment must therefore be done; 2) 
mass loss creates an atmosphere which can extend up to several
tens of stellar radii which renders the use of spherical geometry
unavoidable; 3) the inclusion of metals (mostly iron) is fundamental
to reproduce realistic atmospheric structures and emergent spectra
(line-blanketing effect). 

We are now in an area where powerful
tools including most of the above ingredients with progressively fewer
approximations are becoming available. Examples are the codes TLUSTY
(Hubeny \& Lanz \cite{hl95}), WM-BASIC (Pauldrach et al.\ 
\cite{wmbasic01}) and
FAST-WIND (Santolaya-Rey et al.\ \cite{fast97}). For our study, we have chosen
to use the program CMFGEN (Hillier \& Miller \cite{hm98}) now widely used
for spectroscopic studies of massive stars. The main
ingredients taken into account in CMFGEN are the following: 

\begin{itemize}
\item non-LTE approach: the whole set of statistical equilibrium
equations and radiative transfer equations is solved to yield the
level populations and the radiative field. 

\item wind extension: all equations are written in spherical
geometry with the assumption of spherical symmetry and include
all velocity terms due to the expanding atmosphere.

\item line-blanketing: metals are included in the statistical equilibrium
equations so that accurate level occupation numbers can be derived. A
super-level approximation consisting in gathering level of close
energy in a unique super-level is used to reduce the computational
coast.

\item radiative equilibrium: the temperature structure in the atmosphere
is set by the condition of radiative equilibrium. Other
heating/cooling sources are optional and can be included (adiabatic
cooling, X-rays). 

\item hydrodynamical structure: at present, CMFGEN does not compute the
velocity and density structure so that they have to be taken as input
data or have to be parameterised. In our approach, they are computed
by other atmosphere codes. We use either the ISA-WIND code which uses the
hydrodynamical equations together with a grey LTE temperature to yield 
the density and velocity in the atmosphere (see also Martins et al.\
\cite{teffscale}), or TLUSTY which, thanks to 
a more accurate description of the pressure terms, gives a good
description of the photospheric structure which is connected to a 
wind velocity structure represented by a classical $\beta$
law ($v = v_{\infty}(1-\frac{R_{\star}}{r})^{\beta}$). Tests made with 
stellar and wind parameters typical of
our stars have shown that both methods give similar results for dwarfs
in terms of emergent UV spectra. 
\end{itemize}

Several input parameters have to be specified, the main ones being:

\begin{itemize}
\item stellar parameters: luminosity ($L$), radius ($R$), mass ($M$).

\item wind parameters: mass loss rate ($\dot{M}$, terminal velocity
$v_{\infty}$). Note that the
slope of the velocity field in the wind (the so-called $\beta$
parameter) is chosen when the velocity structure is computed so it is
also an input parameter (with default value 0.8). CMFGEN also gives the 
possibility to include
clumping. This is done by the inclusion of a volume filling factor $f$
of the form 
$f = f_{\infty} + (1-f_{\infty})e^{-\frac{v}{v_{\rm{init}}}}$
where $f_{\infty}$ is the value of $f$ in the outer atmosphere and
$v_{\rm{init}}$ the velocity at which clumping appears (30 km s$^{-1}$
by default in our computations).

\item abundances/elements: in most of our model, the metalicity have been chosen
to be 1/8 solar where solar refers to the values by Grevesse
\& Sauval (\cite{gs98}). This metallicity is thought to be typical of
stars in the SMC and in N81 in particular (Venn \cite{vn99}, Hill
\cite{hill}, Vermeij et al. \cite{vj03}),
although the exact value remains uncertain. 
The ions included in our
computations are given in Table \ref{tab_ions}.

\item turbulent velocity: a value of 20 km s$^{-1}$ was chosen for the
calculation of the populations and of the temperature 
structure. Martins et al. (\cite{teffscale}) have shown that
reasonable changes 
of $v_{\rm{turb}}$ had few effects on this part of the calculation. 
For the formal solution of
the radiative transfer equation leading to the final detailed emergent 
spectrum, the value of $v_{\rm{turb}}$ has been determined to give the best 
fit as shown in Sect.\ \ref{section_vturb} and is found to be of the
order 5 km s$^{-1}$.
\end{itemize}

\begin{table}
\caption{Ions included in the atmosphere models. Number in parenthesis 
  indicate species which, for given \teff, are trace ions and then are 
  not taken into account.}
\label{tab_ions}
\begin{tabular}{ll|llllllll}
\multicolumn{2}{l|}{Element} & & \multicolumn{4}{l}{Ionisation state}  \\
\hline 
H   & & I & II \\
He  & & I & II & III \\
C   & & & (II) & III & IV & V\\
N   & & & (II) & III & IV & V & VI\\
O   & & & (II) & III & IV & V & VI & (VII)\\
Si  & & & (II) & (III) & IV & V\\
S   & & & & (III) & IV & V & VI & (VII)\\
Fe  & & & & (III) & IV & V & VI & VII & (VIII) \\
\hline 
\end{tabular}
\end{table}

\section{Interstellar lines}
\label{interstellar}
The determination of mass loss rates relies on the fit of emission or
P-Cygni lines (see Sect.\ \ref{s_mdot}). The synthetic profiles are
quite sensitive to the value 
of $\dot{M}$, so that a reliable estimate of this quantity requires
the best possible knowledge of the stellar and wind spectrum.
Consequently, the contaminating interstellar (IS) lines must be identified,
which is easy in high resolution spectra where these IS lines appear as
narrow absorptions in the P-Cygni profiles. However, in our
medium resolution observations the IS components are diluted in the
stellar features so that the exact stellar + wind profiles are
uncertain. Of course, this depends on the line: as N~{\sc v} is a trace ion
in the interstellar medium, stellar \nv is weakly affected by this
problem. But this is not the case of \civ which is all the more
contaminated given that the wind feature is weak in the observed N81
spectra. It is therefore crucial to
estimate the contribution of the interstellar absorption to derive
reliable mass loss rates.

The interstellar absorption originates both in the
Galaxy and the local SMC environment. To estimate the interstellar C~{\sc iv}
column densities, we have proceeded as follows:

-- First, we have used high resolution HST-STIS spectra of 8 SMC stars
(AV 69, AV 75, AV 80, AV 327, NGC346 355, NGC346 368, NGC346 113,
NGC346 12), for which UV spectra have been obtained from the HST
archive, 
to determine the C~{\sc iv} column densities in the direction of the NGC 346
region and the southwest part of the SMC. For that purpose, the
Galactic and SMC interstellar profiles have been fitted with a
gaussian profile (with a shift of 140 km s$^{-1}$-- the receding velocity --
for the SMC component). 
Two parameters are needed to achieve such a fit:  
the column density and the FWHM of the gaussian profile
(for which a typical value of 20 km s$^{-1}$ was chosen). The
mean column densities derived from this study are the following:
$\log(N({\rm{C~{\sc IV}}}))=14.28_{-0.13}^{+0.10}$ for the Galactic component, and
$\log(N({\rm{C~{\sc IV}}}))=14.43_{-0.40}^{+0.20}$ for the SMC component. The higher
dispersion in the case of the SMC component is probably due to the
fact that we are looking at different parts of the SMC, whereas only
one line of sight is used for the Galaxy.  

-- Second, we have taken various determination of the C~{\sc iv} column
density from the literature. Several methods were used (curve of
growth, line fitting, apparent optical depth) and give consistent
results.

\begin{table}
\caption{Determination of CIV column densities}
\label{tab_CIV_IS}
\begin{tabular}{lll}
Component & $\log (N({\rm{C~{\sc IV}}}))$ & Reference \\
\hline 
Galactic   & $14.4/14.5_{-0.10}^{+0.10}$  & Mallouris et al. (2001) \\
           & 14.4                         & Fitzpatrick \& Savage (1983) \\
           & $14.35_{-0.06}^{+0.06}$      & Sembach \& Savage (1992) \\
           & $14.28_{-0.13}^{+0.10}$      & SMC stars, this work \\
SMC        & $14.4/14.5_{-0.10}^{+0.10}$  & Mallouris et al. (2001) \\
           & $>$ 14.5                     & Fitzpatrick \& Savage (1983) \\
           & $14.43_{-0.40}^{+0.20}$      & SMC stars, this work \\
\hline 
\end{tabular}
\end{table}

Table \ref{tab_CIV_IS} summarises the various column density estimates.
 For the Galactic case, the results from our determination are
 consistent with more accurate determinations (within the errors
 bars). 
 For subsequent analysis, we adopt {\bf $\log (N({\rm{C~{\sc IV}}})=14.4$} as a
 representative value for the Galactic absorption. For the
 SMC, the results are also in good agreement. 
As the redenning of the SMC-N81 stars is similar to that of the other 8
SMC stars used for this study (the values
of $E(B-V)$ are of the order 0.12 for all stars), the properties of the
interstellar matter on the different lines of sight towards the SMC
sampled here must
be the same. This absence of local extinction was noted by
Heydari-Malayeri et al.\ (\cite{mhm88}) in the first detailed study of
SMC-N81. Then the value of $\log (N({\rm{CIV}})=14.5$ derived on average
for the SMC regions in Table \ref{tab_CIV_IS}
is chosen to be typical of the CIV column density in the
direction of N81.

\begin{figure}
\centerline{\psfig{file=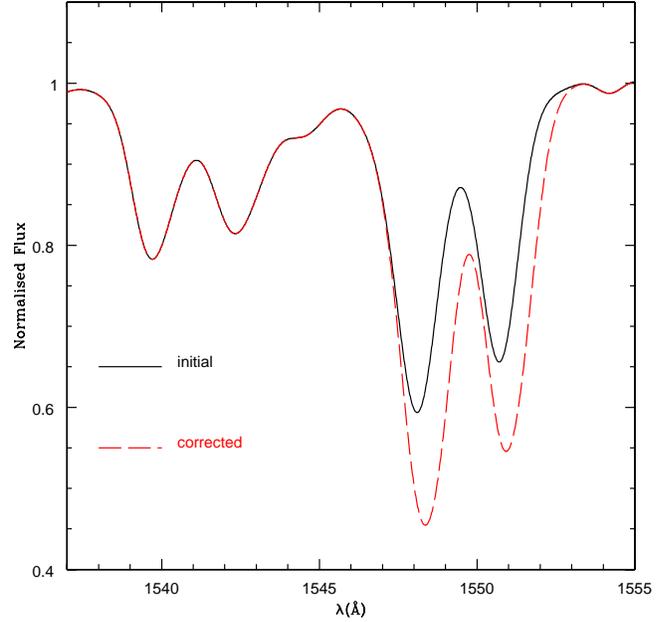,width=9cm}}
\caption{Interstellar C~{\sc IV} component addition: the black solid line
  shows the \civ profile of a CMFGEN model,
  and the
  red long dashed line is the resulting profile after including the 
  IS absorption. A convolution has been performed to take into account
  the instrumental resolution (1.2 \AA). The 
  parameters used to model the interstellar component are: $\log (N(C~{\sc
    IV})_{\rm{gal}})=14.4$, $\log (N(C~{\sc IV})_{\rm{SMC}})=14.5$ and 
  $v_{\rm{SMC}}=140$ km s$^{-1}$.
}
\label{civ_cor}
\end{figure}

With these column densities, we have created synthetic profiles of the
interstellar C~{\sc IV} absorption lines. The method used was the
same as that employed to estimate the column densities from the
spectra of the 8 SMC stars mentioned above. We have then added
these interstellar contributions to the CMFGEN profiles. Figure \ref{civ_cor}
shows an example of such a correction. The typical uncertainties in 
the column densities translate to modifications of the depth of the
corrected absorption profile of the order 0.05.

\section{Detailed analysis of individual stars} 
\label{analysis}
In this section, we present the results of the quantitative analysis of
the UV spectra of our target stars. Constraints
on the effective temperature, the luminosity, the terminal velocity of 
the wind and
the mass loss rate are the main outputs. Secondary constraints on
the slope of the velocity field or the amount of
clumping are also given. The method used is explained in detail for the 
case of star 2, while for the other stars the results are summarised
in Sect.\ \ref{other_stars} and in Table \ref{tab_prop}.

\subsection{Effective temperature}
\label{section_teff}
The most reliable \teff\ diagnostics for O stars remain the
photospheric Helium lines in the optical. Unfortunately, optical
spectra of the SMC-N81 stars are not available so that we had to rely
on the UV. Two types of indicators
were used in this spectral range: the shape of the SED and the
strength of several lines.  

\subsubsection{UV colour index}
\label{colour_index}
As O stars emit most of their luminosity in the 
UV, the shape of the spectral energy distribution at these wavelength 
is sensitive to the
effective temperature just as the optical spectrum in the case of
cooler stars. A colour index in the UV can then
allow to determine \teff. This has been done on the observational side by
Fanelli et al.\ (\cite{fan}) who have computed
various spectral indices based on IUE spectra for different groups of
stars.

\begin{figure}
\centerline{\psfig{file=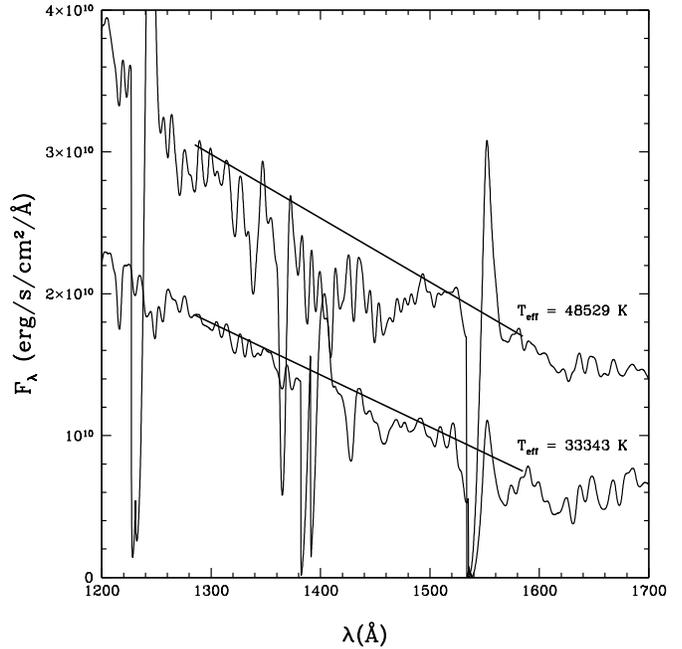,width=9cm}}
\caption{UV flux distribution of two O dwarf models at 33343 K and
  48529 K. Whereas the slope $F_{\rm{1285}}-F_{\rm{1585}}$ increases with
    \teff\, the ratio $\frac{F_{\rm{1285}}}{F_{\rm{1585}}}$ decreases. The bold
    lines are to guide the eye and to show the variations of the 
    slope.
}
\label{fig_UV_slope}
\end{figure}

\begin{figure}
\centerline{\psfig{file=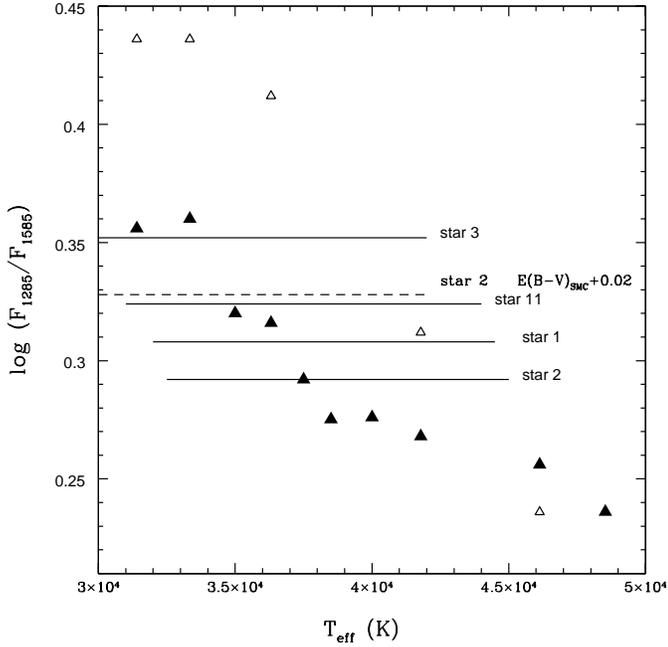,width=9cm}}
\caption{Effective temperature indicator: colour index method. The
  colour index is define by $\frac{F_{\rm{1285}}}{F_{\rm{1585}}}$ where
  $F_{\rm{1285}}$ ($F_{\rm{1585}}$) is the mean flux at 1285 $\AA$ (1585
  $\AA$). Filled (open) triangles are
  for models with
  $Z = 1/8 \zsun$ ($Z = \zsun$). Horizontal lines indicate the colour index for 
  N81 stars which, by comparisons with the theoretical values, give an
  estimate of the effective temperature. Note that this method
  is strongly metallicity-dependent. The dashed line show the position 
  of star 2 for a SMC extinction increased by 0.02.
}
\label{fig_teff_colindex}
\end{figure}

We have used the recent grid of O
dwarf models by Martins et al.\ (\cite{teffscale}) recomputed for an SMC
metallicity ($Z=1/8Z_{\odot})$ completed by various models at this
low $Z$ to derive a relation between
effective temperature and a synthetic colour index defined by
$\frac{F_{\rm{1285}}}{F_{\rm{1585}}}$ where $F_{\rm{1285}}$
($F_{\rm{1585}}$) is the mean flux in an
artificial 20 \AA\ wide  box-shaped filter centered on 1285 \AA\
(1585). The choice of these wavelengths was a compromise between
having fluxes in far enough wavelength ranges (to get a ratio
significantly different from 1) and avoiding metallicity effects
(see below).
In the UV part of the spectrum of interest to us, an increase of
\teff\ translates to a decrease of $\frac{F_{\rm{1285}}}{F_{\rm{1585}}}$
(whereas the slope of the spectrum, $F_{\rm{1285}}-F_{\rm{1585}}$,
increases). This is illustrated in \ Fig. \ \ref{fig_UV_slope} where
we see that an increase of \teff\ from 33343 K to 48529 K induces a
decrease of  $\frac{F_{\rm{1285}}}{F_{\rm{1585}}}$ from $\sim 2.5$ to $\sim
1.8$ while the slope increases (see the bold lines). This is confirmed 
observationally by Fanelli et al.\ (\cite{fan}).

\ Fig. \ \ref{fig_teff_colindex} shows the
correlation between our UV colour index and \teff.  
The determination of the UV colour index from the dereddened flux
distribution of the SMC-N81 stars together with this theoretical
relation allows us to estimate \teff\ of the observed stars.
In the case of star 2, we found a
value of $\sim 40000 K$.

This method suffers from various problems. The most
important is probably the extinction which modifies the slope of the SED
and renders the UV colour index uncertain. The dashed line in \ Fig. \
\ref{fig_teff_colindex} shows the position of star 2 if its SMC
extinction was increased by 0.02 (compared to the extinction derived
from photometry): in
that case, the estimated \teff\ would be  $\sim$ 34000 K. 
This may seem surprising, as naively a higher \teff\ could be expected 
for a higher extinction.
However, this behaviour is simply due to the fact for a ``normal'' 
extinction law (e.g. Pr\'evot et al.\ \cite{prevot}) an increase of 
$E(B-V)$ translates to an increase of
$F_{\rm{dered}}^{\rm{1285}}/F_{\rm{dered}}^{\rm{1585}}$  
which corresponds to a lower \teff\ as discussed above.
Fig.\ \ref{fig_teff_colindex} shows that a typical error of 0.02 on E(B-V) 
(i.e. an error on the flux determination) translates to an uncertainty
of the order 3000/4000 K in \teff\ based on the UV colour index method.

The UV SED is also shaped by many
metallic (mostly iron) lines and thus metallicity can affect the determination of UV 
magnitudes. This is clearly demonstrated in \ Fig.\ 
\ref{fig_teff_colindex} where the effect of increasing the metallicity
from 1/8 \zsun\ to \zsun\ strongly steepens the slope of the relation UV 
colour index - \teff. Both the slope of the continuum and the 
``line forests'' are responsible for such a behaviour.
Our choice of filters centered at 1285 \AA\ and 1585 \AA\
(where the density of metallic lines is weaker than in other wavelength 
ranges) was made to try to minimise the latter effect. 

Finally, the slope of the SED also depends on the gravity (e.g. Abbott 
\& Hummer \cite{ah85}). However,
for dwarf stars, this dependence is weak: a test case run for a model 
with $\log g =$ 4.1 and 4.4 has shown a change of the ratio
$\frac{F_{\rm{1285}}}{F_{\rm{1585}}}$ of less than 2\%.

Given the above uncertainties, the UV colour index method can only give an indication
of the effective temperature which needs to be confirmed by other indicators.

\subsubsection{spectral lines}
\label{teff_lines}

\begin{table}
\caption{Effective temperature estimates.}
\label{tab_teff}
\begin{tabular}{llll}
Star & Estimator & \teff (K) & Adopted \teff (K) \\
\hline 
1   & UV colour index  & 37000 & \\
    & OIV/OV & 36000 & \\
    & Fe~{\sc IV}/Fe~{\sc V} & $\gtrsim$ 38500 & \\
    & N~{\sc V} &  - &  \\
    & C~{\sc III} & 38500 & \\
    &  &  & \bf{38500} \\
2   & UV colour index  & 37500 & \\
    & O~{\sc IV}/O~{\sc V} & $\gtrsim$ 40000 & \\
    & Fe~{\sc IV}/Fe~{\sc V} & 40000 & \\
    & N~{\sc V} & $>$ 35000 &  \\
    & C~{\sc III} & 42000 \\
    &  &  & \bf{40000} \\
3   & UV colour index  & 34000 & \\
    & O~{\sc IV}/O~{\sc V} & $\lesssim$ 37500 & \\
    & Fe~{\sc IV}/Fe~{\sc V} & 36000 & \\
    & N~{\sc V} & $<$ 37500  &  \\
    & C~{\sc III} & 36000 & \\
    &  &  & \bf{36000} \\
11   & UV colour index  & 35000 & \\
    & O~{\sc IV}/O~{\sc V} & 37000 & \\
    & Fe~{\sc IV}/Fe~{\sc V} & $\gtrsim$ 37000 & \\
    & N~{\sc V} & $<$ 37000  &  \\
    & C~{\sc III} & 37000 & \\
    &  &  & \bf{37000} \\
\hline 
\end{tabular}
\end{table}

Several spectral features can be used as \teff\ indicators in the UV. 
\begin{itemize}
\item[$\bullet$] {\oiv/\ov:}
\end{itemize}
    \oiv is present in the spectra of most O dwarfs while \ov appears
    only in stars earlier than O6 (Walborn et al.\ \cite{iue}) so that their
    presence/absence and relative strength is a good \teff\
    estimator. 
    \ Fig. \ \ref{fig_teff} shows comparisons between models of
    different \teff\ and the observed spectrum of star 2. An increase
    of \teff\ increases the strength of both lines which are
    reproduced for \teff\ $\geq$ 40000 K.

    \ov alone has been used by de Koter et al.\ (\cite{dk98}) as 
    a \teff\ indicator for early O stars. Nonetheless, as mentioned by
    these authors, this line is always
    predicted too strong compared to observations above $\sim$ 40000
    K. Below this limit, the situation seems better. Recently, Bouret
    et al.\ 
    (\cite{jc03}) have claimed that clumping could help to solve this 
    well known problem (see Schaerer et al.\ \cite{costar2},
    Pauldrach et al.\ \cite{wmbasic01}): the line is weaker in a
    clumped model than in an homogeneous wind. As \ov\ (together with
    \oiv) depends also on $\dot{M}$, our \teff\ estimate relies on the 
    relative strength of both lines in homogeneous winds. This
    estimate must be confirmed by stronger indicators.

\begin{itemize}
\item[$\bullet$]{Fe~{\sc IV}/Fe~{\sc V}:}
\end{itemize}
 Iron lines forests exist all over the
    UV range. In particular, Fe~{\sc IV} lines are present between 1600 \AA\
    and 1640 \AA, while Fe~{\sc V} lines are found between 1430 \AA\ and 1480
    \AA.  The iron ionisation increases with \teff\ so that Fe~{\sc IV} lines weaken
    relatively to Fe~{\sc V} lines between 35000 and 42000 K. As Fe~{\sc V} is the
    dominant ionisation state of Fe in this temperature range,
    Fe~{\sc V} lines are saturated and little affected by an increase of \teff\
    whereas Fe~{\sc IV} lines weakens. This is 
    shown in Fig.\ \ref{fig_teff}. For star 2, a value
    of at least 40000 K is necessary to reproduce the observed spectrum.

    The determination of \teff\ from this line ratio can be hampered
    mainly by two effects:

    - the iron abundance (and more generally metallicity) can
    change the strength of the Fe absorption. This effect is
    twofold. First, increasing the iron abundance will immediately
    increase the absorption of {\it all} Fe ions, although differently 
    depending on the position of the lines on the curve of
    growth. Consequently, the ratio of lines from two successive
    ionisation states will be modified. 
    Second,
    increasing the iron abundance will strengthen the line-blanketing 
    effect and thus will increase the local temperature in the line
    formation region. The ionisation will be increased,
    leading to a higher \teff\ estimate. The effects of metallicity due 
    to line-blanketing on
    the effective temperature of O stars have been estimated by
    Martins et al.\ 
    (\cite{teffscale}) and turn out to be of the order 1000 to 2000 
    K depending on the spectral type for metallicities ranging from
    solar to 1/8 solar.
    We have run test models for a global metallicity of 
    $Z = 1/5Z_{\odot}$. Its main effect  
    is to strengthen the Fe~{\sc IV} lines, leaving the Fe~{\sc V}
    lines unchanged as they 
    are almost saturated in the temperature range of interest
    here. Quantitatively, this change of Z is 
    equivalent to a decrease of \teff\ by $\sim$ 2000 K as regards the
    iron lines behaviour.

    - the so-called turbulent velocity ''artificially'' strengthens
    the absorption profiles as shown in \ Fig \ \ref{fig_vturb} where
    $v_{\rm{turb}}$  increases from 5 to 15 km s$^{-1}$. Fe~{\sc IV} and
    Fe~{\sc V} lines deepen differently when $v_{\rm{turb}}$ increases,
    so that their relative strength
    remains turbulent dependent. Nonetheless, the effect is smaller
    than the effect on 
    individual lines so that a reasonable estimate of
    \teff\ can be drawn from the study of the relative strength of
    these iron forests. Note that if the effective
    temperature is known, the iron lines can be used to determine the
    turbulent velocity (see Sect.\ \ref{section_vturb}).

\begin{itemize}
\item[$\bullet$]{\ciii:}
\end{itemize}
 As noted by
    Walborn et al. (\cite{iue}) in their IUE atlas of O star spectra,
    this blend of C~{\sc III}
    lines strengthens towards later types. 
    Fig.\ \ref{fig_teff} shows that in the case of star 2 a value of at
    least 42000 K is required to fit the observed spectrum.
    Changing slightly the carbon content can affect this 
    determination. Quantitatively, a reduction of the C abundance from 1/8 C$_{\odot}$ to 1/10 
    C$_{\odot}$ (as discussed in section \ref{mdot_2}) is equivalent
    to an increase of \teff\ by $\sim$ 1500 K.  

\begin{itemize}
\item[$\bullet$]{\nv:}
\end{itemize}
    This resonance line is known to
    be the strongest around spectral type O4 (Smith-Neubig \&
    Bruhweiler \cite{snb97}) and to weaken at later spectral types or
    equivalently when \teff\ decreases. Quantitatively, a \teff\ $>$
    35000 K is
    required to account for the \nv absorption profile
    in star 2. Nonetheless, the strong mass loss dependence of this
    line makes it a poor and only secondary \teff\ indicator: a
    high mass loss associated with a relatively low \teff\ can mimic the 
    N{\sc V} profile of a star with a lower mass loss rate but a higher
    \teff. \\

Table \ref{tab_teff} summarises the results of these \teff\ estimates
which all point to an effective temperature of the 
order 40000 K
for star 2. A real dispersion exists and is mostly due to the
multi-parameter dependence of most of the indicators used.

\subsubsection{Estimation of uncertainty on \teff }
\label{error_teff}

\begin{figure}
\centerline{\psfig{file=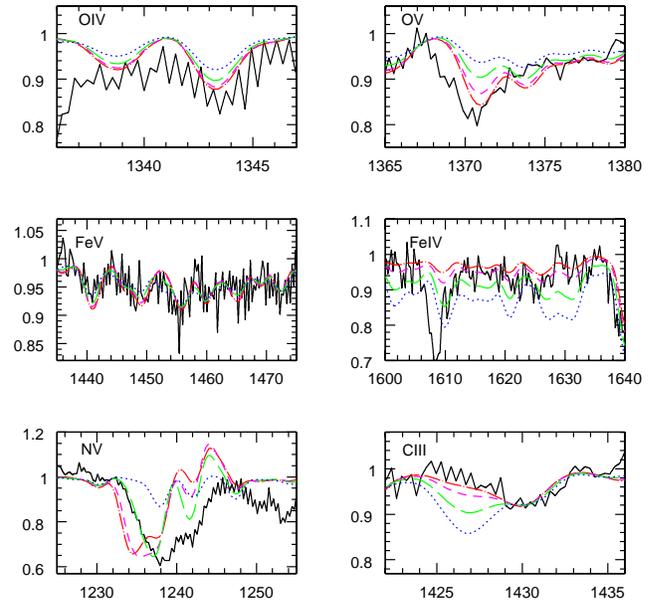,width=9cm}}
\caption{Spectroscopic \teff\ indicators. The black solid line is the
  observed spectra of star 2. Coloured lines are four different
  models with \teff\ 42000 K (red, dot-long dashed line), 40000 K
  (magenta, short dashed line), 37500 K (green, long dashed line),
  35000 K (blue, dotted line). The mass loss rate is fixed at a
  constant value of $10^{-8.5}$ \myr. See text for discussion.
}
\label{fig_teff}
\end{figure}

Table \ref{tab_teff}  shows that for a given star, the dispersion in the
\teff\ estimators is of the order $\pm$ 2500 K in the worst cases. A 
possible additional source of uncertainty comes from the rectification of the spectra. 
We estimate it to be lower than 0.05 in term of normalised flux, which from 
Fig.\ \ref{fig_teff} (especially the Fe{\sc V} plot) can lead to an error on \teff\
of the order $\pm$ 1500 K.

The uncertainties due to variations of the global metallicity and carbon content 
have been estimated above and are of the order 2000 K.

To estimate the efficiency of our \teff\ determination method, we have applied it to 
a star for which optical and UV analysis have given strong constraints on \teff. 
For this purpose, 10Lac was chosen (see section \ref{discussion}). The UV colour index 
method indicate a value of 39000 K, while the spectral lines point to values of the order
35000/36000 K. As the accepted \teff\ for 10Lac is close to 36000 K, the error on the 
\teff\ estimate is not more than $\pm$ 3000 K. This example also shows that 
the spectral lines method is more accurate than the colour index method. For that reason 
we have given more weight to the \teff\ estimates based on the line method in 
our final estimate (as can be shown in Table \ref{tab_teff}).

From the above discussion, we conclude that on average, the uncertainty on our \teff\ 
determination is of the order $\pm$ 3000 K.  \\

\subsection{Mass loss rate}
\label{s_mdot}
The determination of the mass loss rate remains one of the main goals of this
study. As already noted, the spectra of N81 stars show very weak wind
features with no emission. As emission, contrary to absorption, is
entirely formed in the wind, our constraints on
$\dot{M}$ have been set by the requirement that no emission is
produced in the model spectra (as observed in the STIS spectra). Thus, 
we have derived only upper limits on the mass loss rates as models
with $\dot{M}$ below this limit never produce emission. The
main mass loss
indicators in the UV are the \nv, \civ and \ov lines.

\subsubsection{Primary determinations:}
\label{mdot_1}

\begin{figure}
\centerline{\psfig{file=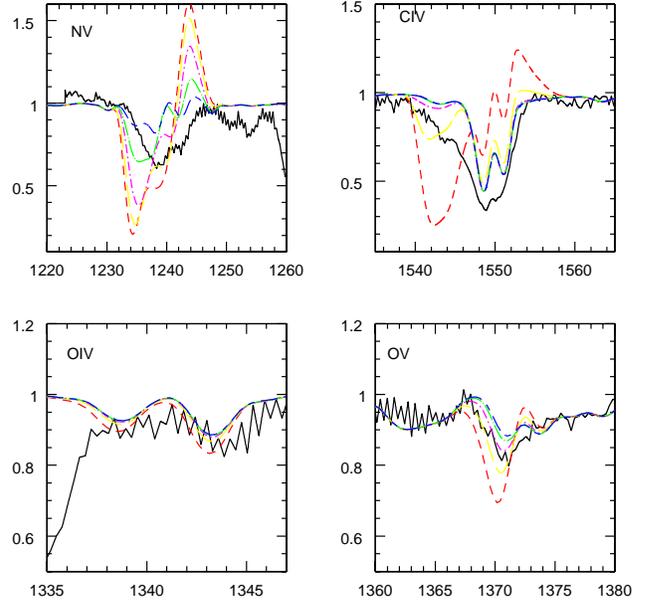,width=9cm}}
\caption{Mass loss rate indicators: the black solid line is the observed
  profile while coloured lines are taken from models with \teff=40000 K 
  and various $\dot{M}$ (-log($\dot{M}$)=7 (red, short dashed line),
  7.5 (yellow, long dashed line), 8 (magenta, dot-short dashed line), 8.5
  (green, dot-long dashed line),  9 (blue, short dashed - long dashed
  line)). See text for discussion.
}
\label{fig_mdot}
\end{figure}

\begin{itemize}
\item[$\bullet$] {\ov:}
\end{itemize}
    \ov develops a P-Cygni profile in the earliest 
    supergiants while only an absorption is seen in dwarfs (Walborn et 
    al.\ \cite{iue}). \ Fig. \
    \ref{fig_mdot} reveals that the profile deepens when $\dot{M} \geq
    10^{-8}$ \myr\ but remains roughly unchanged below this
    value. Fitting the observed spectrum of star 2 requires a mass
    loss rate lower than $10^{-8}$ \myr.
    As mentioned
    in the previous section, \ov is sensitive to \teff\ so that a
    degeneracy $\dot{M}$/\teff\ exists. X-rays can also affect this
    line by increasing the
    Oxygen ionisation to produce O~{\sc VI} (observed in the EUV) by Auger
    ionisation of O~{\sc IV}. 

\begin{itemize}
\item[$\bullet$] {\nv:}
\end{itemize}
 This line is present in dwarfs
    of spectral type earlier than O8 (Walborn et al.\ \cite{iue},
    Smith-Neubig \& Bruhweiler \cite{snb97}) and shows a strong P-Cygni
    profile in early dwarfs. It is mostly formed in the wind so 
    that it is one of the best mass loss 
    rate indicators of the UV part of the spectrum. This is
    illustrated in \ Fig. \ \ref{fig_mdot} where we see the profile 
    changing from a well developed P-Cygni shape to a weak absorption 
    when $\dot{M}$ is reduced by two orders of magnitudes.
    A mass loss rate lower than $10^{-9}$ \myr\ is necessary to 
    produce no emission. However in that case, the absorption profile
    is too weak. Increasing $\dot{M}$ by a factor of 3 improves the
    fit of this absorption part but induces an emission (see Sect.\
    \ref{wind_properties} for a discussion of the shape of these wind
    profiles). Nonetheless, as emission is only produced in the wind
    whereas absorption can originate both in the wind and in the
    photosphere, we prefer adopt $\dot{M} = 10^{-9}$ \myr\ as a 
    reasonable upper limit.

    \nv is also \teff\ sensitive as seen in Sect.\ \ref{teff_lines}
    so that any error on \teff\ can lead to an error on $\dot{M}$. 
    Another problem comes from the X-rays (supposed to be created by
    shocks in the outer wind, e.g. Owocki et al.\ \cite{owo_shock},
    Feldmeier et al.\ \cite{feld}) which can increase
    the ionisation of Nitrogen, thus leading to modifications of the
    \nv profile. A very accurate determination of
    $\dot{M}$ based on \nv would then require the inclusion of X-rays
    in the models. However, tests have revealed that the inclusion of
    X-rays does not modify our conclusions. It is also unlikely that
    X-ray emission is important since any emission in the wind of the
    SMC-N81 stars seems to be very weak as shown by the UV spectra.

\begin{itemize}
\item[$\bullet$] {\civ:} 
\end{itemize}
This line is the
    other strong UV mass loss indicator of 
    O dwarfs. It is seen in all O dwarfs with a
    strength increasing towards early types (Walborn et al.\
    \cite{iue}). In \ Fig.\ \ref{fig_mdot} we see that it
    shifts from a P-Cygni profile to an absorption profile when $\dot{M}$
    decreases. The absence of emission, required by the observation,
    is obtained for $\dot{M} \leq 10^{-7.5}$ \myr. Contrary to
    \nv, a significant absorption profile can remain
    even when the emission disappears. But this absorption is mainly
    photospheric and the wind part turns out to be too weak compared
    to the observed spectra. A change of \teff\ does not lead to any
    improvement (a lower \teff\ implies an emission in \civ\ and a too 
    weak \nv\ line, and a higher \teff\ weakens further the \civ\
    line).\\

\subsubsection{Improving the $\dot{M}$ determinations: effects of
  $\beta$, clumping, adiabatic cooling and abundances}
\label{mdot_2}

   A difficulty to fit the UV lines (especially \civ and \nv) is to
    produce a significant
    absorption extending up to velocities $\geq 1800$ km s$^{-1}$ without any 
    emission. In the following, we concentrate on \civ as the
    discrepancy is highest for this line, but the discussion applies equally
    to \nv.

    The absence or weakness of absorption at high velocity in the
    models showing no emission results from a lack of absorbers
    (i.e. C~{\sc IV} ions) which
    may come either from a too low density, and then from a too low
    mass loss rate, or from a too high ionisation of carbon (CV being
    the dominant ionisation state in the wind). The first possibility
    can be ruled out because a higher mass loss rate will produce a
    non observed emission. Relying on the second hypothesis, we have
    sought for mechanisms that could lead to a reduction of the
    ionisation in the wind. This can be achieved if recombination
    rates are increased, i.e. if the density is higher
    (recombination scales as $\rho^{2}$). As the density is given by 
    $\rho = \frac{\dot{M}}{4 \pi r^{2} v f}$ 
    where f is the filling factor ($f=$1 in an homogeneous atmosphere), a 
    higher density at a given radius can be obtained by either a
    stronger mass loss rate (which is excluded), or a
    clumped wind ($ f<$ 1) or a lower velocity which, as $v_{\infty}$ 
    is fixed, implies a softer
    slope of the velocity field (the  $\beta$
    parameter). Recombinations can also be increased when the
    temperature in the outer wind is reduced. Adiabatic cooling may
    induce such a reduction as in these low density winds it can
    become an important cooling process. Last, abundances can of
    course modify the strength of the wind profiles. In the following, 
    we present investigations of the influence of these various
    parameters on the line profiles and the mass loss rates
    determinations.

\begin{itemize}
\item[$\bullet$] {$\beta$ effects:}
\end{itemize}

\begin{figure}
\centerline{\psfig{file=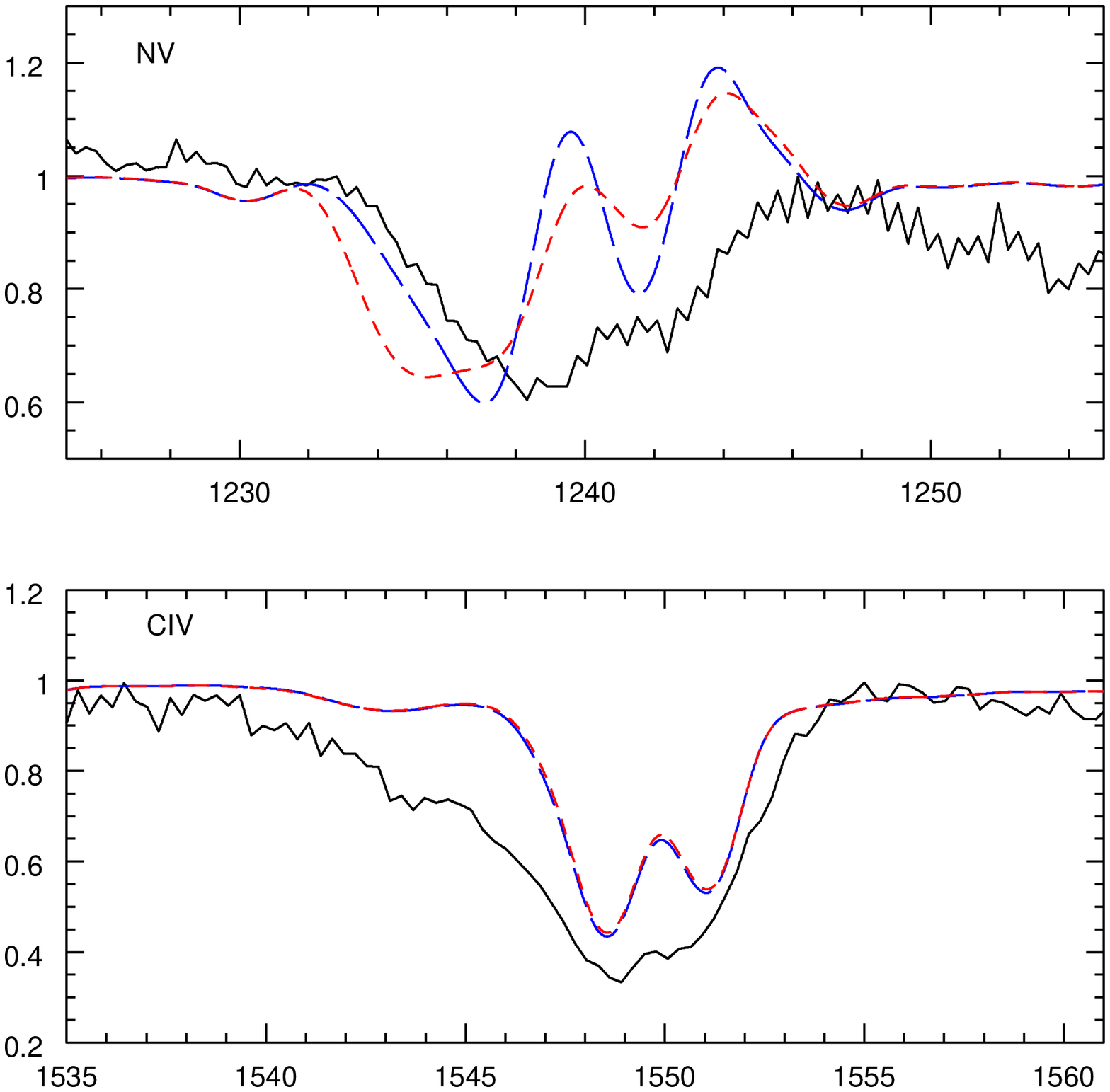,width=9cm}}
\caption{Influence of the slope of the velocity field ($\beta$) on the
  mass loss rate diagnostic lines. Models with $\beta
  = 2.0$ (blue short dashed line) and $\beta =
  0.8$ (red long dashed line) are compared to the observed
  profiles of star 2 (black solid line). Models are for \teff = 40000 K,
  $\dot{M} = 10^{-8.5}$ \myr\ and \vsini\ = 300 km s$^{-1}$. The
  \civ line is almost
  unchanged while the \nv line shows narrower absorption and emission
  when $\beta$ is higher. 
}
\label{beta}
\end{figure}

$\beta$ is usually determined through the shape of Hydrogen emission
lines in the
optical range (e.g. Hillier et al.\ \cite{hil03}). However, as
optical spectra are not available and as the optical lines have probably
absorption profiles due to the weakness of the winds, we have no
constraints on this parameter.
We have then run test models with $\beta = 2.0$ (the 
    default value in all our computations being 0.8). The results are
    displayed in Fig. \ \ref{beta}. \civ is not modified since it is 
    almost purely photospheric in the models shown here, while
    \nv shows narrower and stronger absorption and emission when
    $\beta$ is higher. 
    This behaviour can be explained in term of size of the interaction 
    region which is the region in which, due to the wind velocity
    induced doppler shift, a photon can interact with a given
    line. The size of this region is proportional to the Sobolev
    length (see Lamers \& Cassinelli \cite{lc99}) which scales as
    $(dv/dr)^{-1}$ in the inner wind, so that a higher $\beta$ will
    lead to larger Sobolev length (the acceleration being
    smaller). The interaction region is then wider which leads to a stronger
    absorption or emission in
    the center of the line. In the outer atmosphere, the radial Sobolev
    length is proportional to $\beta^{-1}r^{2}/v$ (for a $\beta$ velocity 
      law) and smaller for higher
    $\beta$ so that the emission/absorption are reduced at high
    velocity
\footnote{When $\beta$ increases, the transverse Sobolev length, which
  is proportional to $(v/r)^{-1}$, increases too. However, the decrease of the radial
  Sobolev length is higher than the increase of the transverse Sobolev 
  length so that globally, the size of the interaction region is
  reduced.}. Illustrations
    of the dependence of the size of the interaction region can be
    seen in Fig.\ 8.6 of Lamers \& Cassinelli (\cite{lc99}).

    In our case, low values of
    $\beta$ seem to be preferred as the observed 
    \nv profile does not show a double absorption line. However,
    $\beta = 2.0$ reproduces better the blue edge of the absorption
    profile. As $\beta =0.8$ is closer to the predictions of the
    radiation driven wind theory, we adopt this value as typical of
    the SMC-N81 stars. But whatever the exact value of $\beta$, the mass loss
    rate determination is not strongly modified as the level of
    emission remains roughly the same (see Fig. \ref{beta}).

\begin{figure}
\centerline{\psfig{file=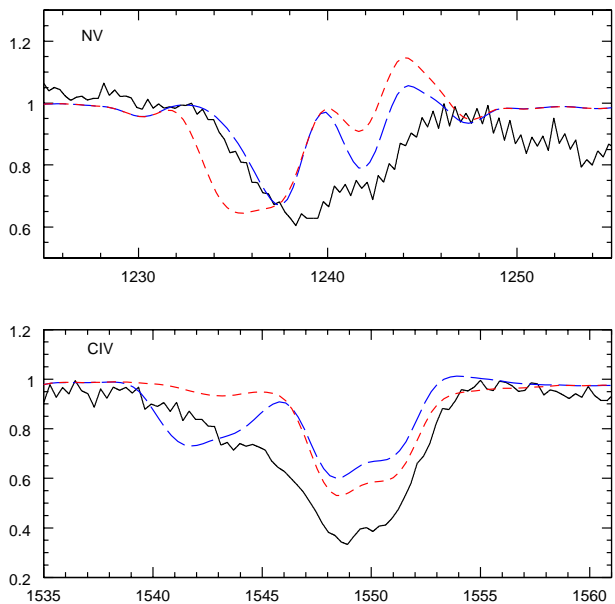,width=9cm}}
\caption{Influence of clumping on the
  mass loss rate diagnostic lines. Models with a volume filling factor 
  of 0.01 (blue long dashed line) and 1.0
  (red short dashed line, no clumping) are compared to the observed
  profiles of star 2 (black solid line). Models are for \teff = 40000 K and
  $\dot{M} = 10^{-8.5}$ \myr\ and \vsini\ = 300 km s$^{-1}$. The
  strength of \nv is
  reduced and the absorption of \civ is increased when clumping is
  included.
}
\label{clumping}
\end{figure}

\begin{itemize}
\item[$\bullet$] {clumping effects:}
\end{itemize}
As already mentioned, there are
    indications that the wind of O stars are clumped (Crowther et al.\
    \cite{paul02}, Hillier et al.\ \cite{hil03}, Bouret et al.\ \cite{jc03}, 
Repolust et al.\ \cite{repolust}), although there exist no
    quantitative constraints. The
    effect of inhomogeneous winds are twofold: first, due to the
    presence of over densities, the emission of density-sensitive
    lines is strengthened; second, the higher density in clumps
    increases the recombination so that the ionisation is reduced. The 
    competition between the two effects can lead to either stronger or 
    weaker emission. To investigate deeper the effect of clumping on
    the wind profiles, we have run
    test models with $f_{\infty} = 0.01$ (recall that the filling
    factor f is given by $f = f_{\infty} +
    (1-f_{\infty})e^{\frac{-v}{v_{\rm{init}}}}$). The main effects on
    the \nv and
    \civ lines are shown in Fig. \ \ref{clumping}. We see that
    the strength of \nv is reduced and that the \civ absorption is
    increased in the outer part of the atmosphere. This is attributed
    to a reduced ionisation in the outer wind (N~{\sc V} recombines to
    N~{\sc IV} 
    and C~{\sc V} to C~{\sc IV}), which improves the fits. Hence,
    there is no doubt that the inclusion of clumping
    is crucial to reproduce the observed features. Whether this is a
    proof of the inhomogeneity of O star winds or just a trick to
    simulate the correct ionisation is not clear, but this parameter
    truns out to be an important ingredient of the modeling. As regards 
    the mass loss determination, Fig.\ \ref{clumping} shows that the
    emission is slightly reduced when clumping is included so that the 
    upper limit on $\dot{M}$ must be slightly increased.

\begin{figure}
\centerline{\psfig{file=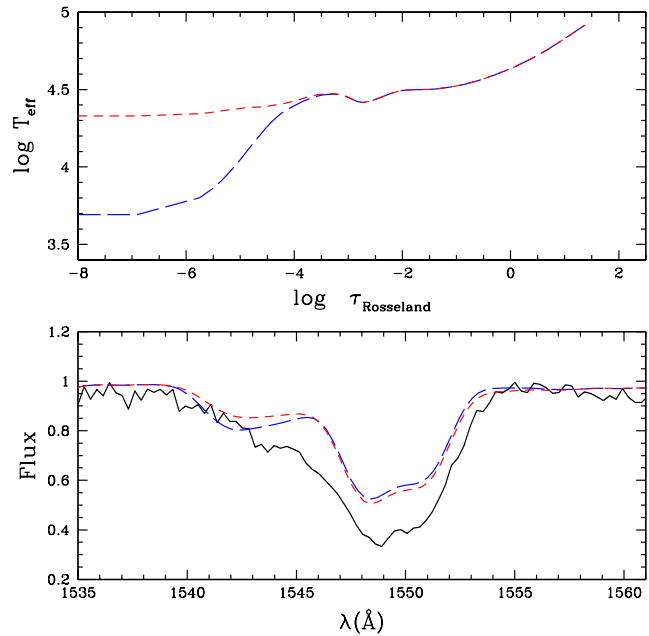,width=9cm}}
\caption{Effect of the inclusion of adiabatic in model computations on 
  the temperature structure (top) and the \civ profile (bottom; blue long dashed
  line: model with adiabatic cooling; red short dashed line: model
  without adiabatic cooling). While the temperature is strongly
  reduced in the outer parts,
  the \civ profile is only slightly affected. Other wind
  lines do not show any change.
}
\label{adcool}
\end{figure}

\begin{itemize}
\item[$\bullet$] {adiabatic cooling:}
\end{itemize}
In low density winds, due to the reduction of any cooling processes
based on atomic mechanisms, adiabatic colling is expected to be
important to set the temperature structure (e.g.\ Drew
\cite{drew}). Fig.\ \ref{adcool}
demonstrates that it is indeed the case: a model with adiabatic
cooling (long dashed line) shows a strong drop of the temperature in
the outer wind. However, as in this part of the atmosphere populations 
are mostly governed by radiative processes, the influence on the line
profiles is reduced: the \civ line shows a slightly enhanced absorption 
which improves the fit but remains marginal. We conclude that
adiabatic cooling is not a crucial parameter \textit{as regards the
  fit of UV wind lines}. 

\begin{itemize}
\item[$\bullet$] {abundances:}
\end{itemize}
Metallicity in our
    models has been chosen to be 1/8 solar, and the individual
    abundances have simply been scaled according to this global
    metallicity. A better assumption would have been to take
    abundances typical of the SMC molecular clouds since  
    the N81 stars are young and their atmospheres probably not contaminated 
    by stellar nucleosynthesis products.
    Such initial abundances can be
    obtained by studies of either main sequence stars which have not yet
    experience internal mixing, or HII
    regions. However, few of the former have been undertaken and are often
    uncertain (Venn 1999 and references therein), and studies of
    nebular abundances can lead to underestimates due to the depletion 
    of material on dust grains. Nonetheless, reasonable agreement
    between these two 
    types of determinations can sometimes be obtained (see Venn \cite{vn99}
    for a discussion) and point to the following values: $C/C_{\odot}
    = 1/10$, $N/N_{\odot}=1/20$ and $O/O_{\odot}=1/5$ (Venn
    \cite{vn99}, Heap \cite{heap03}, Vermeij \cite{vj03}). Adopting
    these abundances in our computations leads to weaker \nv and \civ profiles 
    for a given $\dot{M}$. This means that our upper limit on the mass 
    loss rate necessary to remove emission in the wind lines has to be
    increased. Recently, Asplund \cite{asplund} has revised 
    the solar N abundance. With his new value, the SMC-N81 N abundance derived 
    by Vermeij et al.\ (\cite{vj03}) is 1/30 solar, meaning that the upper limits on $\dot{M}$ 
    has to be increased again a little more.\\

From the above discussion it results that if the CNO abundances are
reduced and set to more realistic values and if clumping is included,
the upper limit on $\dot{M}$ has to be increased by $\sim$ 1.0 dex
(the effect of abundances being dominant). Clumping also helps to get
better shapes of the wind
lines.  $\beta$ and adiabatic cooling have almost no influence on the $\dot{M}$
determination. As a consequence, we adopt an upper limit on the mass loss
rate of star 2 of $10^{-8.0}$ \myr.

\subsection{Reliability of the mass loss rate determination}
\label{test_qi}
In view of the low values of $\dot{M}$ derived, we may wonder whether
our determination is not hampered by any modeling problem. In
particular, what is really determined through fits of UV wind lines is 
the product of the mass loss rate time the ionisation fraction
($q_{\rm{i}}$) of the
absorbing/emitting ion. Any problem with the prediction of these
ionisation fractions would translate in an error on $\dot{M}$. To
investigate this point, two kinds of tests have been pursued.

\begin{itemize}
\item[$\bullet$] {$H_{\alpha}$ vs UV mass loss determination:}
\end{itemize}

$H_{\alpha}$ is much less sensitive to the model
predictions concerning the ionisation than UV resonance lines, so that
we have compared the values of
$\dot{M}$ derived from the fit of UV lines on the one hand and
$H_{\alpha}$ on the other hand. For this aim we have chosen as a test
case the star HD217086 (O7Vn) for which constraints on the mass loss rate
(Puls at al.\ \cite{puls96}, Repolust et al. \ref{repolust}) and 
the ionisation fractions of several ions
(Lamers et al.\ \cite{ion99}) exist. The low density of the wind (see
Lamers et al.\ \cite{ion99}) and the effective temperature of the
order 37000 K make this star similar to the SMC-N81 stars, although
the wind is less weak. We have computed various models to
fit simultaneously
the $H_{\alpha}$ profile and the UV resonance lines.

\begin{figure}
\centerline{\psfig{file=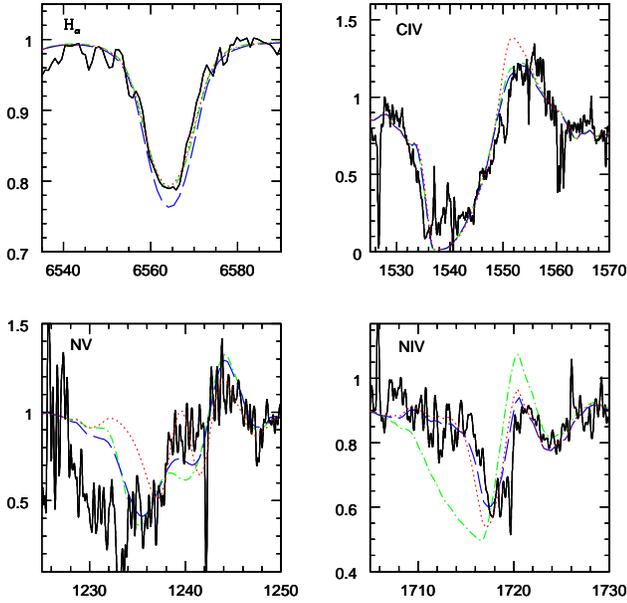,width=9cm}}
\caption{$H_{\alpha}$ versus UV lines $\dot{M}$ determinations. The solid
  line is the observed profiles of wind sensitive lines of
  HD217086. Other lines are CMFGEN models at 37000 K with $\beta =
  1.0$ and  $\dot{M} = 10^{-6.2}$ \myr\ (dot-dashed line),
  $\beta = 1.0$ and $\dot{M} = 10^{-7.2}$ \myr\ (long dashed
  line) and $\beta = 2.0$ and $\dot{M} = 10^{-6.8}$ \myr\
  (dotted line). A rotational velocity of 290 km s$^{-1}$ has been
  adopted. For $\beta = 1.0$ values of $\dot{M}$ different by one
  order of magnitudes are required to fit $H_{\alpha}$ on the one hand 
  and the UV lines on the other hand, while for $\beta = 2.0$ a
  reasonable fit of all lines is achieved. See text for discussion.
}
\label{fig_hd217086}
\end{figure}

Fig.\ \ref{fig_hd217086} shows the results of the fits of the wind
sensitive lines. Two types of 
conclusions can be drawn depending on the value of the $\beta$ parameter
(slope of the wind velocity field):

-- if we adopt $\beta = 0.8$ as derived by Repolust et al. (\cite{repolust}),
the $H_{\alpha}$ profile is best fitted for $\dot{M} = 10^{-6.4}$
\myr\ (dot dashed line) but in that case the UV lines
(especially \niv) are too strong. A value of $\dot{M}$ of
$10^{-7.2}$ \myr\ (long dashed line) has to be adopted to
improve the fit of these 
UV lines, but now the $H_{\alpha}$ absorption is too strong. Mass loss 
rates derived from the UV seem then 
to be lower by a factor 6 compared to the
$H_{\alpha}$ determination. Note that the predictions of Vink et 
al.\ (\cite{vink01}) gives $\dot{M} = 10^{-6.1}$ \myr\ for HD217086.

-- when we increase $\beta$ up to 1.7, a mass loss rate of
$10^{-7.0}$ \myr\  gives a reasonable -- although not perfect --
agreement between observations and models for both $H_{\alpha}$ and
the UV lines (dotted line on Fig.\ \ref{fig_hd217086}).

While the conclusions of the case $\beta = 0.8$ point to a problem
with the ionisation fractions predicted by the CMFGEN models, the case 
$\beta = 1.7$ indicate that this problem partly disappears when $\beta$
is increased. This does not necessarily mean that $\beta = 1.7$ (the
determination of this parameter in low density winds is difficult, see 
e.g. Puls et al.\ \cite{puls96}), but it is a mean to produce an
ionisation throughout the wind which leads to an overall good fit of
the observed profile.
\footnote{As mentioned in Sect.\ \ref{mdot_2} clumping can also modify the
ionisation in the wind. However, several tests have shown that it was
not possible to fit simultaneously $H_{\alpha}$ and UV lines with
clumped models.}
Practically, as we do not have any constraint on $\beta$ (see Sect.\
\ref{mdot_2}) we can conclude that \textit{in the worst case $\dot{M}$ 
  derived from UV lines is underestimated by a \textbf{factor 6}}, at least if 
we take the mass loss rate from $H_{\alpha}$ as the correct one.
In fact, if we take the case $\beta = 1.7$ as representative of the
real conditions in the wind, the error we make when we only look at
the UV spectrum (as for our SMC-N81 stars) and adopt
$\beta = 0.8$ is of 0.2 dex, or less than a factor 2. 
\footnote{Compared to the value 
given by Repolust et al. (\cite{repolust}) - $\dot{M} = 10^{-6.64}$ \myr\ -, 
our best determination ($\dot{M} = 10^{-7.0}$ \myr, $\beta=1.7$) is a factor 
$\sim$ 2 lower}
Moreover, in their recent study of
SMC dwarfs with weak winds, Bouret et al.\ (\cite{jc03}) have been
able to fit
simultaneously lines of different ionisation stages of
the same element with CMFGEN models which indicates that the
wind ionisation was correctly predicted.

\begin{figure}
\centerline{\psfig{file=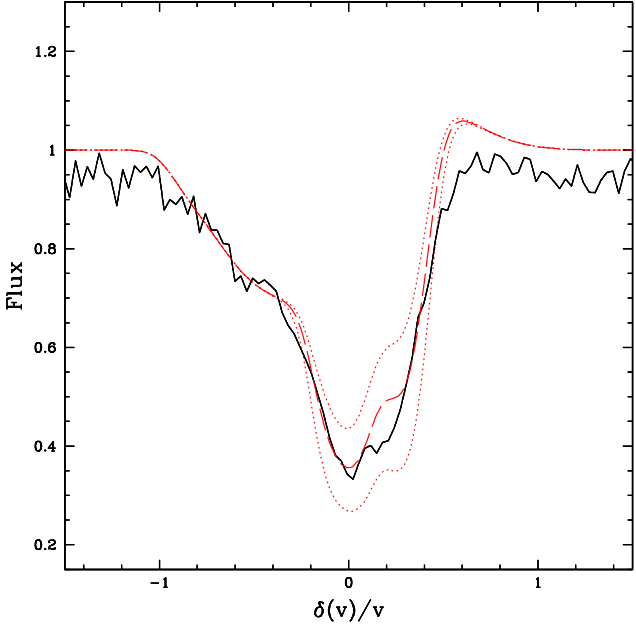,width=9cm}}
\caption{Fit of \civ line with SEI method. The solid line is the observed
  profile and the dashed line is the best fit. The parameters for the
  velocity law are $\beta = 1.0$ and $v_{\infty} = 1800$ km s$^{-1}$. A
  photospheric component has been used. The optical depth has the
  following velocity dependence: $\tau(w) =
  T_{tot}(1-(w)^{1/\beta})(w)^{0.2}$ with $w =
  v/v_{\infty}$ and $T_{tot} = \int_{0.01}^{1} \tau(w) $. The derived
  value of $\dot{M} \times q_{\rm{i}}$ is $10^{-9.68}$. Dotted lines show the
  influence of an increased or reduced photospheric component: the wind part of the
  absorption is not affected by this component.
}
\label{sei}
\end{figure}

\begin{itemize}
\item[$\bullet$] {SEI method:}
\end{itemize}
We have applied the SEI method (Lamers et al \cite{sei87}) to our SMC N81
star 2. This method leads to the determination of $\dot{M} \times q_{\rm{i}}$. Basically,
the SEI method solves the radiative transfer in an expanding
atmosphere for which the source functions are calculated in the
Sobolev approximation. The main input parameters are the velocity
field and a function giving the optical depth of the line as a
function of velocity. The main output is the value of $\dot{M} \times q_{\rm{i}}$. 
The best fit to the \civ line of star 2 is shown in \ Fig. \
\ref{sei}. The corresponding value of $\log (\dot{M} \times q_{\rm{i}})$ is -9.68.
We also show in this figure the influence of the underlying
photospheric absorption: as expected, the low-velocity part of the \civ
absorption profile is sensitive to this photospheric
absorption. As the interstellar contamination renders the true
stellar absorption uncertain (see above),
we do not give too much weight to this part of the profile
which can always be fitted by tuning the photospheric absorption. The
important point is that the wind profile ($0.4 \leq \Delta
v/v_{\infty} \leq 1.0$) is insensitive to the amount of photospheric
absorption and only depends on the wind parameters. Our estimation of
$\dot{M} \times q_{\rm{i}}$ with the SEI method can then be regarded as reliable.

In order to compare this value to that deduced from our model giving the 
best fit for star 2, we have computed the ionisation fraction of
C~{\sc IV}
using the definition of Lamers et al.\ (\cite{ion99}):

\begin{equation}
q_{\rm{i}} = \frac{\int_{x_{\rm{0}}}^{x_{\rm{1}}}n_{\rm{i}}(x)
  dx}{\int_{x_{\rm{0}}}^{x_{\rm{1}}}n_{\rm{E}}(x) dx}
\end{equation}

where $n_{\rm{i}}$ is the
population of the absorbing ion, $n_{\rm{E}}$ the population of the
element, $x_{\rm{0}}$ ($x_{\rm{1}}$) the lower (higher) integration limit ($x$
being $r/R_{\star}$). We found $\log {q_{\rm{C~{\sc IV}}}}  = -1.36$ which, together
with a mass loss rate
of $10^{-8.5}$ \myr\ gives $\log(\dot{M}q_{\rm{i}})$ = -9.86, in
reasonable agreement with the SEI result.
This shows that the value of $\dot{M} \times q_{\rm{i}}$ obtained by
fitting the CMFGEN model profiles to the observations is correct. Thus 
it follows that if $\dot{M}$ is underestimated, 
the C~{\sc IV} ionisation fraction
is overestimated. To investigate this
point, we have compared our ionisation fractions to those derived by
Lamers et al.\ (\cite{ion99}). For stars of different stellar and wind 
properties, they found a mean $q_{\rm{C~{\sc IV}}}$ of the order $10^{-2.5}$ which
is $\sim$ a factor of 10 lower than the CMFGEN value. An error by
such a factor on $ q_{\rm{i}}$ translates to an underestimate of $\dot{M}$ by the 
same factor. This confirms the result of the previous section where a
possible underestimate of $\dot{M}$ by a factor 6 was highlighted.

An interesting comment to make is that if we simply \textit{assume}
that the derived mass loss rate is correct, the mean density in the
wind (as defined by Lamers et al.\ \cite{ion99}) is of the order
$10^{-16...-17}$ g cm$^{-3}$ which is well below the lowest density
probed by the Lamers et al.\ sample (see their Fig. 3). As their
analysis indicates an increase of the C~{\sc IV} ionisation fraction with
decreasing mean density, it is conceivable that for very weak winds,
$q_{\rm{C~{\sc IV}}}$ may reach values of the order 0.1, which would
reconcile the
CMFGEN and Lamers et al.\ ionisation fractions. Moreover, it should be
noted that the study of Lamers et al.\ includes radio and $H_{\alpha}$ 
mass loss rates which did not take clumping into account. Since the
inclusion of clumping leads to lower $\dot{M}$ (e.g. Hillier et al.\
\cite{hil03}), it is conceivable that some of the Lamers et al.\
ionisation fractions are underestimated.\\

The main conclusion of these two types of studies (H$_{\alpha}$/UV and SEI)
is that there are indications that the ionisation
fractions predicted by CMFGEN may be wrong, but by no more then a factor 
of $\sim$ 10 (leading in that case to an underestimation of $\dot{M}$ by the
same factor). However, in both studies we have also found means to
explain the observations with the CMFGEN predictions (changing the
value of $\beta$ in the study of HD217086, extrapolating the trend
$q_{\rm{C~{\sc IV}}}$ - mean density in the SEI study). It is thus not clear if the
ionisation fraction predictions of the CMFGEN models are erroneous or
not, especially given that we are in a range of parameters never explored
before (weak winds). However, if these predictions are wrong, they do not
\textit{qualitatively} modify the conclusion concerning the weakness
of the winds (see Sect.\ \ref{wind_properties}).

\subsection{Terminal velocity}
\label{section_vinfty}
\begin{figure}
\centerline{\psfig{file=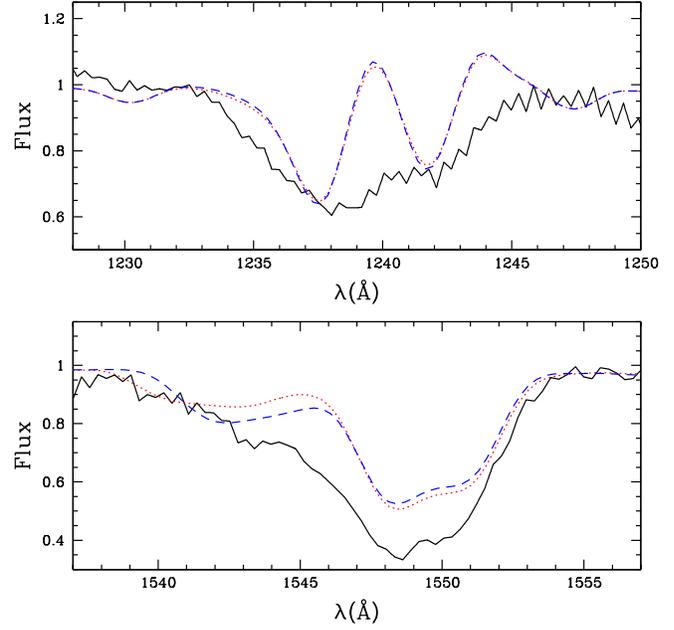,width=9cm}}
\caption{Determination of $v_{\infty}$. The solid line is the observed 
  spectrum while the dotted (dashed) line is a model with
  $v_{\infty}=1800 (1500)$ km s$^{-1}$. No change is seen in \nv and 
  \civ is hardly affected. See text for discussion.
}
\label{fig_vinf}
\end{figure}

  The terminal velocity of the winds of O stars is usually derived
  from the blueward extension of UV resonance lines. Here, due to the
  weakness of the outer wind density, the absorption may not extend up 
  to $v_{\infty}$ so that with the above method one can only derive
  lower limits for the terminal velocities. 
  As shown in
  Table \ref{tab_prop}, these limits can even be lower than the
  escape velocity (which is of the order 1100 km s$^{-1}$),
  reinforcing the fact that the absorption probably does not extend up 
  to $v_{\infty}$.
  Moreover, the relatively low signal
  to noise ratio of our spectra coupled to the uncertainty in the flux 
  normalisation introduces an uncertainty in the exact position of 
  the most blue-shifted absorption.
  Figure \ref{fig_vinf} shows the \nv and \civ profiles 
  of models with $v_{\infty} = 1500$ and $1800$ km s$^{-1}$ compared to the
  observed line. The \nv profile is not affected by the change of
  $v_{\infty}$ while the \civ profile is hardly modified. In view of
  this result and as the fit
  with the SEI method requires a value of 1800 km s$^{-1}$ for the terminal
  velocity, we adopt this value as representative for star 2. The
  values for the other stars are given in Table \ref{tab_prop}.

\subsection{Turbulent velocity / Rotational velocity}
\label{section_vturb}
\begin{figure}
\centerline{\psfig{file=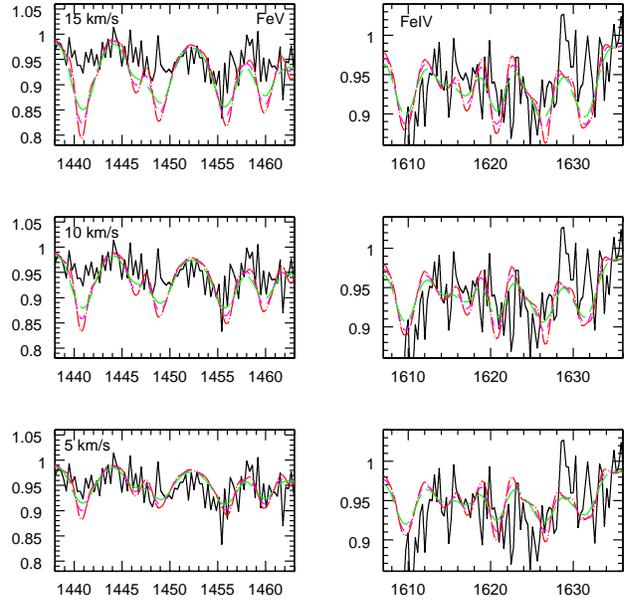,width=9cm}}
\caption{Effect of microturbulence and rotation on the iron spectrum. 
 The three left
  panels show Fe~{\sc V} lines while the Fe~{\sc IV} lines are shown on the right
  panels. The turbulent velocity decreases from 15 to 5 km s$^{-1}$ from top
  to bottom. Each model is convolved with a rotational velocity of 100 
  (dot - long dashed line), 200 (short dashed line) and 300 km s$^{-1}$ (long dashed
  line). The lower panels ($v_{\rm{turb}} = 5$ km s$^{-1}$) give the best fit.
}
\label{fig_vturb}
\end{figure}

The turbulent velocity in O stars is thought to increase from values
of a few km s$^{-1}$ near the photosphere to $\sim 10 \%$ of the
terminal velocity in the outer atmosphere. 
As mentioned in Sect.\
\ref{section_teff}, iron lines are sensitive to $v_{\rm{turb}}$ so that
once \teff\ is known, they can be used to determine the turbulent
velocity. \ Fig. \ \ref{fig_vturb} shows the comparison of
iron lines of a model with $\teff = 40000$ K but different
$v_{\rm{turb}}$. For each model, spectra convolved with rotational
velocity of 200, 250 and 300 km s$^{-1}$ are shown. The best fit is
obtained for turbulent velocity of 5 km s$^{-1}$. For higher values, the Fe~{\sc V}
lines deepens too much compared to the observed spectrum. The Fe~{\sc IV}
lines behave similarly but also weakly. Tests have been run with a
turbulent velocity varying from 5 km s$^{-1}$ near the photosphere up to 100
km s$^{-1}$ in the outer wind and have revealed very little changes in the
blue part of wind profiles. This is explained by the weakness of the
wind in which absorption is likely to take place up to velocities
lower than the terminal velocity of the wind.
\ Fig. \ \ref{fig_vturb} also reveals that the rotational velocity of
the star is of the order 300 km s$^{-1}$. Indeed, lower velocities lead
to narrow and deep profiles not observed.

\subsection{Luminosity}
\label{luminosity}
The luminosity of the stars has been estimated from the absolute
visual magnitude $M_{\rm{V}}$ and a bolometric correction calculated for the
estimated \teff\ of the star according to 

\begin{equation}
\log \frac{L}{L_\odot}=-0.4 (M_{\rm{V}}+BC-M^{bol}_{\odot})
\end{equation}

where $M^{bol}_{\odot}=4.75$ (Allen \cite{allen}). $M_{V}$ was derived in paper I from the
observed visual magnitude and the estimated extinction. BC, which is
essentially model independent when calculated as a function of \teff, is
derived from the relation of Vacca et al.\ (\cite{vacca})

\begin{equation}
BC(\teff)=27.66-6.84 \times \log \teff
\end{equation}

The uncertainty is 0.01 for the observed visual magnitude (paper I),
0.05 for the extinction (from an uncertainty of 0.01 on $E(B-V)$), 0.025 
for the distance modulus (di Benedetto \cite{dibene}) and finally 0.25 
for $BC$ for a typical error on \teff\ of 3000 K. On average, the
uncertainty on $L$ is therefore of the order 0.12 dex.

\subsection{Summary and results for other stars}
\label{other_stars}
Stellar and wind parameters for 3 other stars of SMC-N81 have been
determined with a similar analysis. Here we summarise the results of
this analysis and give the results in Table \ref{tab_prop}. The best
fits are shown in the Appendix. For the
remaining stars presented in paper I,
no constraints have been derived due to the poor quality of the
spectra. For all the stars, $\beta$ = 0.8 and Si, S and Fe
abundances equal to 1/8 solar have been chosen. The masses 
have been derived from the HR diagram presented in Fig.\
\ref{HR_diag}. 

\subsubsection{star 2}
The main results have been given in the previous sections. 
In contrast to all other stars, we stress
that the fits of the \civ\ and \nv\ lines for star 2 are improved when the
abundances are taken 1/8 solar. Choosing the lower values indicated by 
the nebular analysis (see Sect.\ \ref{mdot_2}) lead to too weak
absorptions even when clumping is included. The best fits are achieved 
with a filling factor at the top of the atmosphere of 0.01
\footnote{Recall
that in the present parameterisation of clumping in CMFGEN, this means 
that $f$ goes from 1 at the photosphere to 0.01 when $v = v_{\infty}$.}.

\subsubsection{star 1}
For this star, \teff\ is found to be $\sim$ 38500 K. The upper limit on 
the mass loss rate is 10$^{-8.5}$ \myr\ and $v_{\infty}$ is at least
  1500 km s$^{-1}$. Clumping is necessary to improve the fit of the
  \civ\ line ($f_{\infty}$ = 0.01). A reasonable fit is achieved with
  the nebular CNO abundances. 

\subsubsection{star 3}
Star 3 is the coolest of the 4 stars studied (\teff\ = 36000 K). The
determination of the mass loss rate is very difficult because the \nv\ 
line is almost absent due to the low \teff\ and the \civ\ line is
almost entirely photospheric. Hence, secondary indicators such as
\oiv\ have been used estimate $\dot{M}$ (which is found to be lower
than  10$^{-8.5}$ \myr). We want to stress that this
determination is probably the most uncertain of our sample, 
especially since the
the observed spectrum is noisy. The very small value for the 
terminal velocity reflects the low density of the wind: the number of
absorbant in the wind is so low that the wind lines are essentially
absent. The nebular CNO abundances give the best fits. 

\subsubsection{star 11}
We have estimated an effective temperature of 37000 K for this star. 
An upper limit of 10$^{-9.5}$ \myr\ is estimated from the \nv\ and
  \civ\ lines. The terminal velocity we derive (600 km s$^{-1}$) is
  again a lower limit. The nebular CNO abundances give reasonable fits 
  and clumping is not necessary.

\begin{table*}
\caption{Summary of the stellar and wind properties derived for the N81 
  stars. The spectral types are estimated from the optical spectra of
  the models giving the best UV fits. As we have only lower limits on
  the terminal velocities, we have adopted v$_{\infty}$/v$_{esc}$ =
  2.6 to compute both the modified wind momenta and the theoretical
  mass loss rates of 
  Vink et al.\ (\cite{vink01}). The Si, S anf Fe abundances are 1/8
  solar and n(He)/n(H) is 0.1. The gravity adopted for our 
  computation was log g = 4.0 as it is typical of dwarf 
  stars (Vacca et al. \cite{vacca}) and as we have no diagnostics to estimate 
  the value of this parameter.}
\label{tab_prop}
\center
\begin{tabular}{lll|lllllllllllll}
 & & & & & star1 & & & star2 & & & star3 & & & star11 &  \\
\hline 
 & & & & & & & & & & & & & & &\\
$m_{\rm{V}}$& & & & & 14.38 & & & 14.87 & & & 16.10 & & & 15.74 & \\
E(B-V)& & & & & 0.07 & & & 0.06 & & & 0.10 & & & 0.07 & \\
 & & & & & & & & & & & & & & &\\
Spectral type & & & & & O7 & & & O6.5 & & & O8.5 & & & O7.5 & \\
$M_{\rm{V}}$& & & & & -4.84 & & & -4.32 & & & -3.21 & & & -3.48 & \\
\teff [K]& & & & & 38500 & & & 40000 & & & 36000 & & & 37000 & \\
$\log (L/L_{\odot})$& & & & & 5.32 & & & 5.16 & & & 4.59 & & & 4.73 & \\
$R/R_{\odot}$& & & & & 10.3 & & & 7.9 & & & 5.0 & & & 6.9 & \\
$M/M_{\odot}$& & & & & 32 & & & 30 & & & 19 & & & 21 & \\
\vsini\ [km s$^{-1}$]& & & & & 200 & & & 300 & & & 250 & & & 250 & \\
$\log \dot{M}$ [\msun yr$^{-1}$]& & & & & $\lesssim$ -8.0 & & & $\lesssim$ -8.0
& & & $\lesssim$ -8.5 & & & $\lesssim$ -9.0 & \\
$v_{\infty}$ [km s$^{-1}$]& & & & & $\geq$1500 & & & $\geq$1800 & & &
$\geq$300 & & & $\geq$ 600 & \\
$v_{esc}$& & & & & 1088 & & & 1203 & & & 1204 & & & 1077 & \\
$log (\dot{M} v_{\infty} \sqrt{R})$& & & & & $\lesssim$ 26.76 & & &
$\lesssim$ 26.75 & & & $\lesssim$ 26.14 & & & $\lesssim$ 25.67& \\
$f_{\infty}$& & & & & 0.01 & & & 0.01 & & & 1 & & & 1 & \\
$\log Q_{0}$ [s$^{-1}$]& & & & & 48.85 & & & 48.76 & & & 47.99 & & & 48.41 & \\
$\log \dot{M}_{\rm{Vink}}$ [\msun yr$^{-1}$]& & & & & -6.96 & & & -7.25 & & & -8.31
& & & -8.04 & \\
 & & & & & & & & & & & & & & &\\
C/C$_{\odot}$ & & & & & 1/10 & & & 1/8 & & & 1/10 & & & 1/10 & \\
N/N$_{\odot}$ & & & & & 1/20 & & & 1/8 & & & 1/20 & & & 1/20 & \\
O/O$_{\odot}$ & & & & & 1/5 & & & 1/8 & & & 1/5 & & & 1/5 & \\
\hline 
\end{tabular}
\end{table*}

\section{Nebular and stellar properties} 
\label{neb_ste_prop}
In this section we first go
back to the nebular properties of N81 and then investigate the
evolutionary status of the individual stars together with the
consequences of their weak winds.

\subsection{Nebular properties / Ionising fluxes}
A meaningful way of testing our results concerning the stellar
properties of the N81 individual components is to compare them to the
integrated properties of the cluster. 

First, Heydari-Malayeri et al.\ (\cite{pap0}) derived a mean extinction 
of $A_{V} = 0.40$ from the observed $H_{\alpha}/H_{\beta}$
ratio in the nebula, and they use this value to correct the observed
flux in $H_{\beta}$ and to estimate the number of Lyman continuum 
photons - $Q_{\rm{0}}$ - emitted by the N81 stars (under the assumption that the HII
region is ionisation bounded). They find $Q_{\rm{0}} = 1.36 \times 10^{49}$
photons $s^{-1}$. From this and the calibration of Vacca et
al.\ (\cite{vacca}), they conclude that a single main sequence star of
spectral type O6.5 or O7 can lead to such an ionising flux. 
We have estimated the total amount of ionising photons released
by the N81 stars from the SEDs of the models giving the best UV fit. As 
the N81 stars studied here are the most luminous and hottest of the
region, they are likely to provide essentially all the ionising flux. We
found from the models $Q_{\rm{0}} = 1.64 \times 10^{49}$ photons $s^{-1}$
in good agreement with the
values derived from the nebular properties. The lower value of the
estimate of Heydari-Malayeri et al.\ (\cite{pap0}) can be partly
explained by the fact that the HII region may be density bounded so
that a part of the ionising flux may escape the cavity. 

\begin{figure}
\centerline{\psfig{file=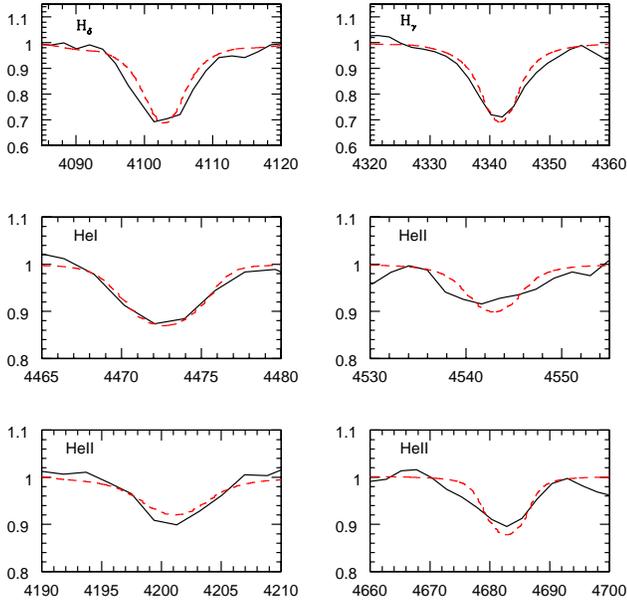,width=9cm}}
\caption{Comparison between observed optical spectrum of the N81
  cluster corrected for nebular lines (solid line) and sum of the
  optical spectra of the models giving the best UV fits (dashed
  line). Each optical
  spectra from the models has been convolved with rotational
  velocities derived from the UV spectra. The agreement is very good,
  showing that our determination of the stellar parameters of the N81
  stars is reliable.
}
\label{fig_comp_neb}
\end{figure}

Second, we have integrated optical spectra of the N81 cluster that have
been corrected for the nebular contamination so that they give the total
stellar spectrum which can be compared to the sum of the individual
stellar optical spectra of the models giving the best
UV fits. Fig.\ \ref{fig_comp_neb} shows that
there is again a good agreement between the observed and modeled
spectra. From the strength of the He lines, a ``mean'' spectral type
O7 is found (which is in
fact similar to the spectral type of the two most luminous stars of
the cluster, namely star 1 and 2).
This indicates
that despite the difficulties in estimating the stellar properties of
the individual stars (in particular \teff), our results are
reliable. Moreover, the He~{\sc ii}  $\lambda$4686 line is reasonably
fitted by the models. As this line is usually filled by wind emission, 
this may be an indication of not too strong mass loss rates.

\subsection{Evolutionary status}
\begin{figure}
\centerline{\psfig{file=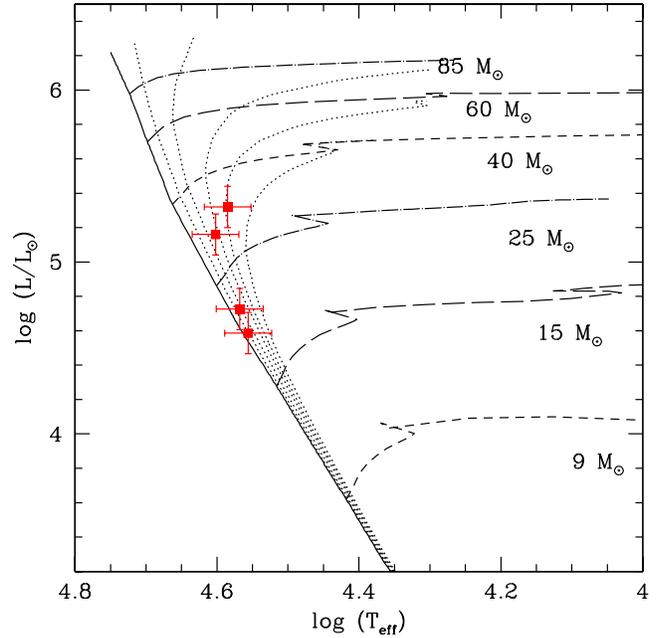,width=9cm}}
\caption{HR diagram for $Z=0.004$ without rotation.
 Different Geneva evolutionary tracks for various masses are indicated,
 together with the ZAMS and isochrones for 1,2,3,4 and 5 Myrs (data from 
 Lejeune \& Schaerer \cite{lej01}). The
 filled squares give the position of the SMC-N81 stars with the
 typical errors on their position. The calibration $\log L - log 
 \teff\ $ of Vacca et al.\ (\cite{vacca}) for dwarfs is also shown by the 
 long dashed line. Note the underluminosity of most of the SMC-N81 stars.
}
\label{HR_diag}
\end{figure}

We have placed the N81 stars on an HR diagram constructed with Geneva
evolutionary tracks without rotation for$ Z=0.004$ (Fig.\
\ref{HR_diag}). 
All the stars are compatible with an age $\leq$ 5 Myrs. While star 1
seems to be slightly older, the 3 other stars (especially star 3 and
11) lie close to the ZAMS and have an age between 0 and 4 Myrs. There
seems to be an age dispersion of the order 1 or 2 Myrs in the cluster, 
which is reasonable.
The inclusion of
rotation is known to move the ZAMS towards lower \teff\ (see Meynet \& 
Maeder \cite{mm97}), reducing the age estimates for stellar
populations compared to the non-rotating case. However, this effect
becomes significant only for rotational velocities close to the
critical velocity. As the SMC-N81 stars have \vsini\ of
only 200-300 km s$^{-1}$ (which corresponds to $\sim$ 1/3 of the break
velocity), our results are probably not strongly hampered by
the use of stellar tracks without rotation.

Based on 
the equations governing the dynamical evolution of HII regions (Dyson
\cite{dyson}) with
the typical values $\dot{M}=3 \times 10^{-9}$ \myr\,
$v_{\infty}=1500$ km s$^{-1}$, $n_{\rm{0}}=400$ cm$^{-3}$ (gas density, see
Heydari-Malayeri et al.\ \cite{mhm88}) and a radius of the HII region
of 3 pc, one can derive an age of $\sim$ 1.4 Myrs for the N81
region. This assumes that star 1 and 2 provide most of the mechanical
energy ($\frac{1}{2}\dot{M}v_{\infty}$). 
Previous age estimates based on
the decrement of $H_{\beta}$ equivalent width by Heydari-Malayeri et al.\
(\cite{mhm88}) were of 1 to 2.5 Myrs assuming a stellar cluster
of solar metallicity.
However, for clusters with a small number of ionising stars this 
method is inherently inaccurate due to statistical fluctuations.

If we compare the positions of the SMC-N81 stars with the calibration
$L$ - \teff\ for dwarfs of Vacca et al.\ (\cite{vacca}) based on studies of
Galactic  O stars (long dashed line in Fig.\ \ref{HR_diag}), we note that
except star 1, the other components are all underluminous compared to
``normal'' O dwarfs. This result based on the quantitative study of
the stellar properties confirms the conclusion of paper I in which the 
sub luminosity of most of the stars was already highlighted 
and interpreted as an indication of the youth 
of the stars.

In paper I it was argued that the SMC-N81 stars could belong to
the class of Vz stars. Fig.\ \ref{HR_diag} shows that although
young, the stars do not lie perfectly on the ZAMS. This
may be an indication that either Vz stars are not strictly ZAMS stars but
more probably ``young'' stars less evolved than ``normal'' dwarfs, or
that our objects are not 
``true'' Vz stars (we recall that we do not
dispose of optical spectroscopy to firmly establish if the SMC-N81
stars are Vz stars or not). A deeper
investigation of the properties of this interesting class of objects
will be presented in a forthcoming paper.

\section{Puzzling wind properties}
\label{wind_properties}

\subsection{Comparison to previous observations and theoretical
  predictions}

One of the most surprising feature of the spectra of the SMC-N81 stars 
is the shape of the wind lines (mainly \nv and \civ): while they show
no emission, they display a blueshifted absorption profile. 
This is quite puzzling: how can we produce such an absorption without any
emission ? Theoretically, in an extended atmosphere, the absorption in the
P-Cygni profile comes from the removal of photons from the observer's line of
sight, and the emission is due to photons that have been scattered
isotropically on 
sight lines not parallel to the observer's one. Globally, the amount
of ``absorbed'' photons is then equal to the number of ``emitted''
photons. If the atmosphere is only weakly extended (i.e. its height
is smaller than the stellar radius), then roughly half the
photons are backscattered towards the star and destroyed, so that the
emission is reduced compared to the absorption (the ratio between them 
being 1/2). This is even more true if there is an underlying
photospheric absorption which increases the absorption with respect to 
the emission. This points to the direction observed in our spectra,
but the absence of emission remains a puzzle.

A possible explanation for the absence of emission may be an enhanced
backscattering of 
photons towards the photosphere, so that the emission is even more
reduced compared to the absorption. But the physical reason for
such a process is not known. Another possibility is a strongly non
spherically symmetric wind. Rotation is known to increase the mass loss 
near the poles and to reduce it at the equator (Bjorkmann \&
Cassinelli \cite{bc93}, Maeder \& Meynet
\cite{mm00}). If a star rotates sufficiently fast to induce a strong
contrast between the polar and equatorial ejection, and if we can
observe it pole-on, then we can obtain the kind of profile we have:
the blueshifted absorption comes from the high density polar flow
while the (absence of) emission corresponds to the low density ``equatorial''
atmosphere. However, as \textit{all} the SMC-N81 stars show such
profiles and as the probability to see all of them pole-on is low,
this explanation is not satisfying.
Moreover, as the projected rotational velocities of the N81 stars
are of the order 200/300 km s$^{-1}$, it is not likely that they are seen pole-on.

\begin{figure}
\centerline{\psfig{file=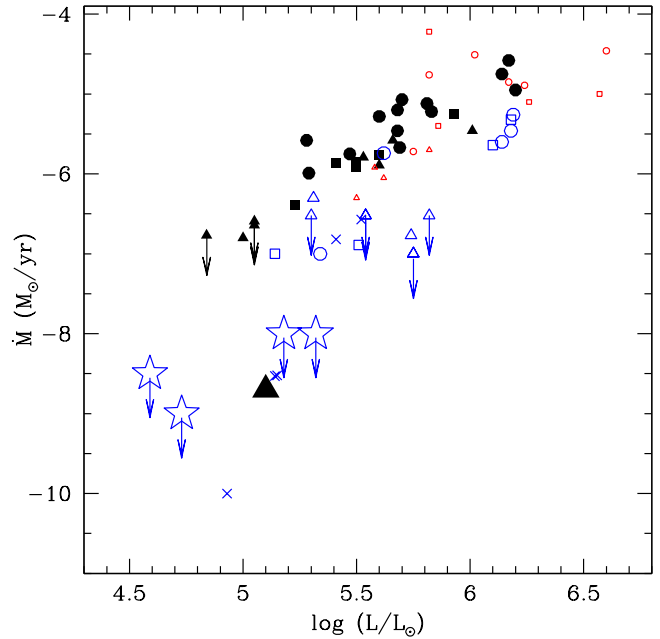,width=9cm}}
\caption{Mass loss rate as a function of stellar luminosity for O stars. 
  Filled (open) symbols are Galactic (LMC, small red / SMC, blue)
  objects. Triangles 
  (squares, circles) are for luminosity class V (III, I). Crosses are
  the SMC stars of Bouret et al.\ (\cite{jc03}). Data are
  from Puls et al.\ (\cite{puls96}), Herrero et al. (\cite{hpv00}),
  Crowther et al. (\cite{paul02}),
  Repolust et al.\ (\cite{repolust}), 
  Hillier et al. (\cite{hil03}) and Massey et al. (\cite{massey04}). 
  The star symbols are for the SMC-N81
  stars. Note the low mass loss rates of
  the SMC objects with $\log \frac{L}{L_\odot} \leq 5.5$ and of 10 Lac 
  (large filled triangle).
}
\label{fig_mdot_L}
\end{figure}

In spite of this curious features, we have been able to place
constraints on the wind properties of the SMC-N81 stars, the main results
being that they lose mass at an extremely low rate. Whether
this is typical or not of O dwarfs at low metallicity 
remains to be established.
Indeed, this is one of the first quantitative
determination of mass loss rates for such objects and consequently few 
comparisons are possible. To our knowledge, the only previous studies
of wind parameters O dwarfs in the SMC have been performed by Puls et
al.\ (\cite{puls96}), Bouret et al.\ (\cite{jc03}) and very recently
Massey et al.\ (\cite{massey04}). Puls et al.\
(\cite{puls96}) have only derived upper limits on the mass loss rates
for four SMC dwarfs. This is mainly due to the fact that their
method relies on the the strength of the $H_{\alpha}$ emission whereas 
the $H_{\alpha}$ profile in SMC dwarfs is mostly in absorption. Their
upper limits are of the order $10^{-7}$ \myr\ which is roughly
10 times higher than our estimates for the SMC-N81 stars. The recent
work by Bouret et al.\ (\cite{jc03}) include 5 SMC O dwarfs of which
three (with spectral type between O9.5 and O6) have mass loss rates 
between $10^{-10}$ and $3 \times 10^{-9}$
\myr. Our results agree very well with these estimates
(we recall that the SMC-N81 stars have mid to late spectral
types). The dwarfs studied by Massey et al.\ (\cite{massey04}) are
more luminous than the N81 stars and have therefore higher mass loss
rates. Fig.\ \ref{fig_mdot_L} shows mass loss rates as a function of
luminosity for Galactic and SMC dwarfs from different studies. The
results of this work and of Bouret et al.\ (\cite{jc03}) indicate a
strong reduction of the mass loss rate of stars with $\log
\frac{L}{L_{\odot}} \lesssim 5.5$.

How do these results compare to theoretical calculations ? The most
recent predictions of wind parameters as a function of
metallicity are the extensive calculations by Vink, de Koter \& Lamers
(\cite{vink01}). Using their cooking recipe to estimate the mass loss 
rate as a function of stellar parameters and metallicity, we found
$\dot{M}$ of the order $10^{-7..8}$ \myr\ for the SMC-N81 stars
(the exact values are given in Table \ref{tab_prop}). The derived mass 
loss rate are more than 30 times lower. Even if we take into account a
possible error in the ionisation fraction of the CMFGEN models (which
would lead to an increase 
of the $\dot{M}$ determinations by a factor of $\la$ 10) the mass
loss rates are still lower than predicted (with perhaps the exception
of star 3 for which we recall that the $\dot{M}$ determination is
uncertain). There is thus no doubt that the winds of the
SMC-N81 stars show an unusual weakness.

\begin{figure}
\centerline{\psfig{file=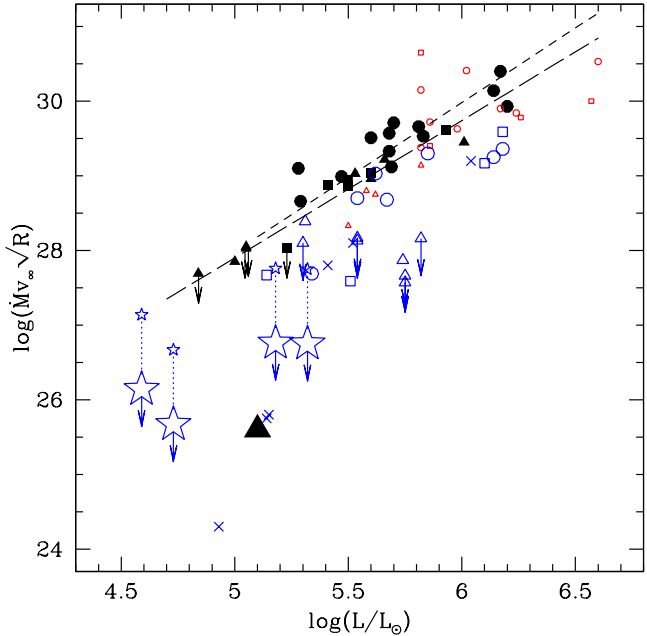,width=9cm}}
\caption{Modified wind momentum - luminosity relation. Data and
  symbols are the same as in Fig.\ \ref{fig_mdot_L}. The SMC-N81 stars 
  (for which we have adopted v$_{\infty}$=2.6 v$_{esc}$ to compute the 
  modified wind momentum)
  confirm the trend of either a reduced modified wind momentum at
  low luminosity or a steeper slope of this relation at low
  metallicity. The small star symbols show the position of the SMC-N81 
  stars if $\dot{M}$ was systematically underestimated by a factor 10. The long -
  dashed (short - dashed) line gives the mean
  relation for luminosity clas V - III (I) stars from Repolust et
  al. (\cite{repolust}). The position of 10 Lac is also indicated by the 
  large filled triangle.
}
\label{wlr}
\end{figure}

This result is even more striking in term of modified wind
momentum. This quantity defined by $\dot{M} v_{\infty}
\sqrt R$ is predicted to be a power law of the sole stellar 
luminosity (e.g. Kudritzki \& Puls.\ \cite{kp00}) to give the so-called 
modified wind momentum - luminosity relation (hereafter WLR)

\begin{equation}
\dot{M} v_{\infty} \sqrt R \propto L^{\frac{1}{\alpha}}
\label{eq:eq_wlr}
\end{equation}

\ Fig. \ref{wlr} shows the
observed relation for Galactic and MC stars. For Galactic stars, the
correlation is well defined and is slightly different for luminosity class
I stars and LC class III-V stars (as demonstrated by the regression
curves). 
For the MC objects however, the relation seem to be different from
that followed by the Galactic objects, at least for the LC V
stars. Indeed, our results indicate that the SMC-N81 stars
(star symbols if \ Fig.\ \ref{wlr}) have 
modified wind momenta lower by $\sim$ 2 orders of magnitudes compared
to Galactic objects of the same luminosity and following the LC III-V
relation, which simply reflects the weakness of the winds
\footnote{To compute the modified wind momenta of the SMC-N81 stars,
  we have adopted v$_{\infty}$ = 2.6 v$_{esc}$ since we have not been
  able to derive the true terminal velocities due to the weakness of
  the winds.}.
The study of Bouret et al.\ (\cite{jc03}) indicate a similar 
trend. Note that even if the mass loss rates are underestimated by a factor
10 (see Sect.\ \ref{mdot_2}) there is still a significant difference
compared to Galactic objects (except in the case of star 3) as shown
by the small star symbols in Fig. \
\ref{wlr}. We have also compared the derived wind momenta with those
expected for B and A supergiants (e.g. Kudritzki et al.\
\cite{kud99}): it turns out that the SMC-N81 momenta are weaker. This
confirms the extreme weakness of the winds of the SMC-N81 stars.

\subsection{Possible origin of the low $\dot{M}$/wind momentum}
\label{discussion}

Several possibilities can be invoked to
explain the low wind momenta observed for the SMC-N81 stars:
first, there may be a breakdown of the modified wind momentum -
luminosity relation at low luminosity for dwarfs; 
second, metallicity can affect
this relation; third, ion runaway may happen in the wind, reducing the 
mass loss rate. In the following, we investigate these different hypothesis.
Finally we will speculate on the possible relation between these
``unusual'' wind properties and the youth of these stars.\\

1) \textit{Breakdown of the wind momentum - luminosity
relation at low luminosities:}

Let us first consider the case of Galactic stars.
The study of Puls
et al.\ (\cite{puls96}) showed that in the luminosity range $5.2 < \log
L/L_{\odot} < 5.5$ several of their stars seemed to indicate a weaker
modified wind momentum than expected from the relation followed by
higher luminosity stars indicating a possible break in the slope of
the WLR. However, the recent reanalysis of 
these stars by Repolust et al.\ (\cite{repolust}) do not show this trend 
any more 
if their upper limits are taken as the true values for $\dot{M}$.
As their mass loss determinations are based on
$H_{\alpha}$ fitting and as this line becomes insensitive to $\dot{M}$ 
below $\sim 10^{-8}$ \myr\ (e.g. Herrero et al.\ \cite{hpn02}), 
one can not rule out the possibility of low wind momenta for these low 
luminosity objects. UV analysis of these stars could certainly shed
more light on this question. 

Concerning low metallicity objects, few LMC stars have been studied
so far, all of them being furthermore of high luminosity.
More SMC stars have been studied (including the
SMC-N81 objects). For objects with $log L/L_{\odot} > 5.5$,
the wind momenta are lower than for Galactic stars, but the WLR
seems to have the same slope. The low luminosity stars show a clear
reduction of the wind momenta which imply a steeper slope of the WLR
in this luminosity range. Nonetheless, most of the high L objects are
giants or supergiants while low $L$ objects are all dwarfs. As the WLR
may be different for different luminosity classes, the existence of a
real breakdown is unclear. 

Moreover, reasons for such a possible break down of the wind momentum -
luminosity relation are not known. Is it linked with the
driving mechanism ? A possibility could be a change of the ions
responsible for the radiative acceleration with the consequence of the 
modification of the efficiency of the driving, but this would be
linked to the change of the ionisation in the atmosphere and then more 
to \teff\ than to $L$. Is this break down of the wind momentum
luminosity relation only due to the low luminosity of the stars ?
There are in fact other objects, central stars of planetary nebulae,
which are much less luminous than any O stars and which seem to follow 
the mean relation for Galactic objects with $\log \frac{L}{L_\odot}
\geq 5.5$ (see Kudritzki \& Puls \cite{kp00}). This renders the
behaviour of the low luminosity O stars even more puzzling.

Another possibility is that the current formalism of the radiation
theory may be erroneous in the case of massive stars with weak
winds. Indeed, Owocki \& Puls (\cite{op99}) have shown that in such
winds, the curvature of the velocity field near the sonic point could
lead to an \textit{inward} directed flux of the diffuse radiation
field that can significantly reduce the total radiative
acceleration compared to the CAK approach (see their Fig.\ 7). As a
consequence, the mass loss rates are also reduced compared to the
predictions based on the CAK formalism for the radiative
acceleration. For stars with high luminosity and/or low gravity, the
density scale height just above the photosphere is higher 
\footnote{The density scale height is
proportional to $R_{\star}^{2}/M(1-\Gamma)$ where $R_{\star}$ is the
stellar radius and $\Gamma$ the ratio of electron scattering to
gravitational acceleration} so that line thermalisation near the sonic 
point may suppress the above effect (see Owocki \& Puls
\cite{op99}). This may explain why giants/supergiants and dwarfs
with high luminosity are not sensitive to the effects of the curved
velocity field and show a more classical behaviour. 
This possibility is attractive since it could explain why
only dwarfs with low luminosities seem to have wind properties
deviating from the predictions of the CAK theory.

Puls, Springmann \& Lennon (\cite{psl00}) have made a detailed
study of the line statistics and its effect on the radiative driving
and have shown that under 
certain conditions, the parameterisation of the radiative acceleration 
with the CAK formalism ($g \propto (\frac{dv}{dr})^{\alpha}$) is not
valid, and then the predictions based on this formalism are
erroneous. This would happen if both the level density of the ions
responsible for the driving were much higher \textit{and} the
distribution of the oscillator strengths were much steeper. However,
they argue that the atomic physics of the driving ions should not lead to
such extreme conditions. 

In summary,
the above discussion presents evidence that the current predictions of 
the radiation driven wind theory may fail to explain the winds of low
luminosity O dwarfs due to subtle effects negligible in the winds of
stars studied so far.\\

2) \textit{Metallicity effect:}

\begin{figure}
\centerline{\psfig{file=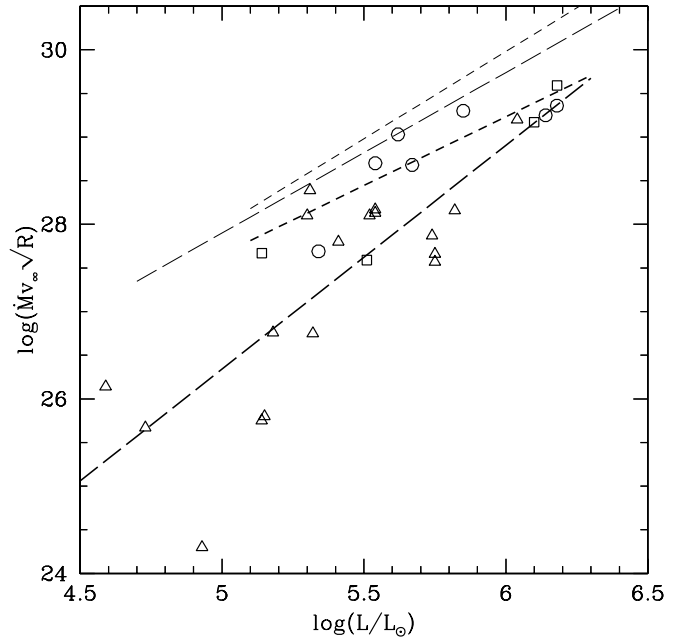,width=9cm}}
\caption{Modified wind momentum - luminosity relation for SMC
  stars. Data are from Puls et al.\ (\cite{puls96}), Bouret et al.\
  (\cite{jc03}), Massey et al.\ (\cite{massey04}) and the present
  work. Dwarfs (giants, supergiants) are shown by triangles (squares,
  circles). The long - dashed (short - dashed) curves are regressions
  for supergiants (giants/dwarfs) of the Galaxy (light curves, from
  Repolust et al.\ \cite{repolust})) and the SMC (bold curves). SMC
  giants and dwarfs may indicate a steeper slope of the modified wind
  momentum - relation at low metallicity, but the scatter is important
  and more studies are needed to confirm this trend.
}
\label{wlr_smc}
\end{figure}

The wind momentum - luminosity relation may be different at low
metallicity. Such a dependence, though quite weak, is in fact
predicted by the radiation driven wind
theory. The inverse of the luminosity exponent in equation
\ref{eq:eq_wlr} ($\alpha$)
is linked to the line strength distribution function. But this 
function depends on metallicity, $\alpha$ being lower at low $Z$
(e.g. Abbott \cite{abbott82},  Puls, Springmann \& Lennon \cite{psl00}). It turns 
out that the
wind momentum - luminosity relation should have a steeper slope at sub 
solar metallicities. The reduction of the number of lines effectively
driving the wind also leads to a global shift of the relation towards
lower modified wind momenta. Our results tend to indicate
a steeper slope of the WLR,
together with the Puls et al.\ (\cite{puls96}) SMC objects and the
Bouret et al.\ (\cite{jc03}) results. Nonetheless, the real form of
the WLR is

\begin{equation}
\dot{M} v_{\infty} \sqrt R \propto
(M(1-\Gamma))^{\frac{3}{2}-\frac{1}{\alpha}} L^{\frac{1}{\alpha}}
\end{equation}
\label{eq_wlr_m}

where $\Gamma$ is the ratio of radiative acceleration to
gravitational acceleration.
For solar metallicity, $\alpha \sim 2/3$ so that the dependence on the 
effective mass $M(1-\Gamma)$ vanishes. Values of $\alpha < 2/3$
brings back this dependence, which should lead to a large scatter of
the WLR for low $Z$ objects and then to a more difficult calibration of
this relation. Fig.\ \ref{wlr_smc} shows the modified wind
  momentum - luminosity relation for the SMC stars studied so far: the
  scatter is indeed important. Indicative regression curves are given
  and reveal a possible steeper slope ($\alpha \sim$ 0.4) for
  giants/dwarfs while for the supergiants $\alpha$ may be close to
  2/3. More studies are needed to confirm this trend of a 
steeper slope at lower metallicities.

However, two points may impede this explanation. First, 
the most recent hydrodynamical simulations do 
not show any change of the slope of the WLR at metallicities typical
of the Magellanic Clouds (Vink et al.\ \cite{vink01}, Hoffmann et al.\ 
\cite{tadziu02}). Furthermore, Kudritzki (\cite{kud02}) has shown that 
only for
extremely low metal content -- $10^{-3} Z_{\odot}$ -- a difference
appears, although his calculations were made
at high luminosities ($\log L/L_{\odot} > 6$) were most of the
driving is thought to be done by H and He lines, leading to a weaker
metallicity dependence of the wind properties. Second, some Galactic
stars are known to display signatures of weaker than normal winds 
(Walborn et al.\ \cite{iue}) but none has been analysed
quantitatively so far. 

\begin{figure}
\centerline{\psfig{file=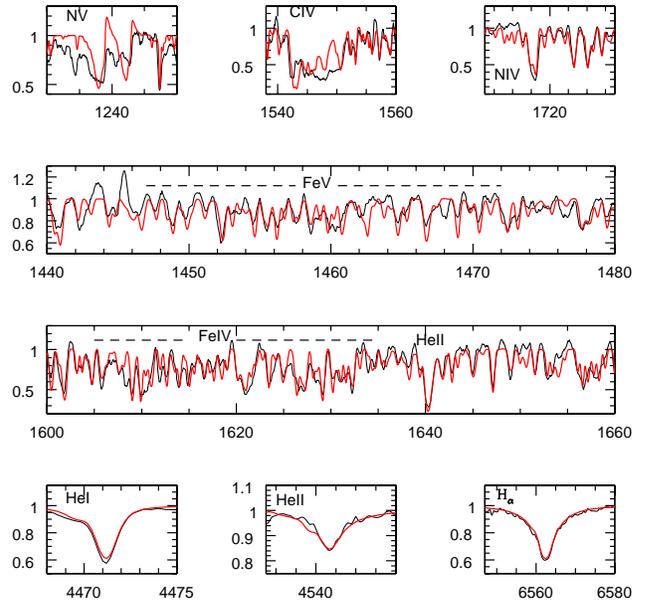,width=9cm}}
\caption{Fit of the UV and optical spectra of 10 Lac (light curve:
  observation; bold curve: model). The parameters of the
  models are: $\teff = 36000$ K, $\log g = 4.1$, $\dot{M} = 2 \times
  10^{-9}$\myr\, $v_{\infty} = 1070$ km s$^{-1}$. A turbulent velocity
  increasing from 5 km s$^{-1}$ near the photosphere to 100 km s$^{-1}$ in the outer 
  wind has been adopted. Abundances are from Herrero et al.\
  (\cite{hpn02}) for He, Villamariz et
  al. (\cite{vil02}) for CNO and have been set to the solar value (Grevesse \& 
  Sauval \cite{gs98}) for the other metals. A rotational velocity of
  40 km s$^{-1}$ has been adopted.
}
\label{10lac}
\end{figure}

We have carried out a detailed 
study of the wind properties of one of them: 10 Lac.
This O9V star has often been considered as a
standard dwarf star for its stellar properties. We
have run models to fit its IUE and optical spectra taking the 
the stellar parameters from Herrero et al.\ (\cite{hpn02})
\footnote{\teff $=$ 36000 K, $\log g =3.95$, 
and the He abundance $\epsilon = 0.09$.} 
and the CNO abundances from Villamariz et al.\ (\cite{vil02}).
A microturbulent velocity
increasing from 5 km s$^{-1}$ near the photosphere up to 100 km
s$^{-1}$ in the outer atmosphere was used. 
Fig.\ \ref{10lac} shows the
results of our best fit model for which $\dot{M} = 2 \times 10^{-9}$
\myr\ and $v_{\infty} = 1070$ km s$^{-1}$. This value of $\dot{M}$ is
lower than 
estimates of Howarth \& Prinja (\cite{hp89}, $10^{-6.7}$ \myr\ based
on line profile fitting with a Sobolev code), of Leitherer 
(\cite{claus88}, $10^{-6.83}$\myr\ based on $H_{\alpha}$) and lower
than the prediction of Vink et al.\ (\cite{vink01}) - $10^{-6.2}$ \myr\ 
-
but is in agreement with the more recent upper limit of
Herrero et al.\ (\cite{hpn02}) based on $H_{\alpha}$ fits ($10^{-8}$
\myr). The important point is that the wind momentum of this
star is as low as for the SMC-N81 stars (see Fig.\ \ref{wlr}). 
The
metal content of 10 Lac being near solar (10 Lac belongs to the 
nearby association Lac OB1, and its CNO abundances are
between 0.5 and 0.9 the solar values - see Villamariz et al.\
\cite{vil02}), this 
indicates that metallicity is not uniquely responsible for the
weakness of the winds observed in the present work.

The above discussion shows that our understanding of the metallicity
effects on the wind properties of massive stars is still partial and
that both observational and theoretical efforts are needed to improve
this situation.\\

3) \textit{Decoupling:} 

Finally, an alternative answer to the puzzling behaviour of the wind
properties of the low luminosity O stars is that a decoupling between
the absorbing ions and the passive plasma may occur. In classical
radiatively driven winds, ions from metals gain momentum from the
photons they absorb and redistribute this momentum to the passive
plasma through frictions. However, if the density in low enough, these 
frictions may become inefficient so that the absorbing ions are the
only species to be accelerated. Due to their low abundances compared
to H and He (constituting most of the passive plasma), the mass loss
is greatly reduced.

Various studies have been pursued to determine the conditions under 
which such decoupling may take place (e.g Babel \cite{babel95},
\cite{babel96}, Krti\v{c}ka \& Kub\'{a}t \cite{kk00}). Springmann \&
Pauldrach (\cite{sp92}) have estimated that decoupling occur when
radiative acceleration is not compensated by frictional braking. We
have applied their equation 15 to the case of our SMC-N81 stars and
we have found that decoupling is expected to happen only in the outer
region of the wind where the outflow velocity has almost reached the
terminal velocity. In the transsonic region, where the mass loss rate 
is set, decoupling is not predicted to take place. More
recently, Owocki \& Puls (\cite{op00}) have given an estimate of the wind
velocity at which maximal coupling occur as a function of stellar and
wind parameters. If this velocity is greater than the terminal
velocity, then the wind can be described by a single component
fluid. In the SMC-N81 stars, this maximal drift is
predicted to happen at velocities of the order
$v_{\infty}$ so that decoupling is not expected to happen. Even more
recently, Krti\v{c}ka et al.\ (\cite{krti03}) have estimated the
metallicity below which decoupling should occur due to the
reduced wind density for a given set of stellar parameters. In the
case of the SMC-N81 stars, this limit is of the
order $1/100 Z_{\odot}$ which is well below the metallicity of the SMC. 
As a consequence, it seems that decoupling cannot explain the
weakness of the winds.\\

After these considerations the most likely explanation for the weak
winds of the analysed stars is therefore that a breakdown of the 
modified wind momentum - luminosity relation exists at low luminosity 
for dwarfs (1), possibly independently of metallicity.
A possible mechanism responsible for such a behaviour may exist 
(cf.\ above) although no realistic calculations have been done
for parameters appropriate to the objects considered here.
In any case, it is not clear what parameter(s) would determine
that some O dwarf stars have such weak winds in comparison
to other O stars of similar luminosity.
Possible relevant parameters could be a higher gravity, 
higher mass/light ratio, or others (such as the presence of magnetic
fields). 
From the current limited sample of objects showing such
puzzling wind properties we speculate that the low mass loss rate
is probably intrinsically related to the youth of the stars.
It is conceivable that we are beginning to witness the
``onset'' of radiatively driven winds in young, still somewhat
underluminous O stars shortly after their formation.
Further systematic studies of Vz stars and related objects with 
indications of weak winds will be necessary to resolve these issues.

\section{Conclusion} 
\label{conclusion}
Based on UV spectral obtained with STIS/HST 
we have analysed the stellar and wind properties of the four main exciting
stars of the High Excitation Blob SMC-N81 using extensive
calculations of spherically expanding non-LTE line blanketed 
atmosphere models with the code CMFGEN. 

The main results are the following:

\begin{itemize}
\item[$\diamond$] {The stellar properties (L, \teff) indicate that
    the SMC-N81 components are young ($\sim$ 0--4 Myrs old) O stars which 
    shows, with perhaps the exception of star 1, a
    lower luminosity than ``normal'' Galactic O dwarfs. This,
    together with the closeness to the ZAMS for star 3 and 11,
    confirms the conclusion of paper I that they may belong to the Vz
    class (Walborn \& Parker \cite{wp92}). }

\item[$\diamond$] {The UV spectra of the N81 stars show unusually weak 
    stellar winds. The upper limits on mass loss rates are of the
    order a few
    $10^{-9}$ \myr\ which is low compared to 1) Galactic stars
    of the same luminosity and 2) the most recent predictions of
    $\dot{M}$ as a function of stellar parameters and
    metallicity. Point 1) could be qualitatively understood due to the
    reduced metallicity of the SMC but point 2) indicates that this
    reduction is higher than expected.}

    Although the mass loss rates derived from the UV line analysis
    are potentially affected by uncertainties in the modeled ionisation
    fractions, various tests indicate that the above conclusions remain 
    qualitatively valid.

\item[$\diamond$] {Our objects show modified wind momenta 
    ($M_{\odot} v_{\infty} R^{1/2}$)
    which are, for the same luminosity $L$, lower by typically 
    two orders of magnitude compared to the ``normal'' O star samples.
    Similarly low wind momenta have also been found by Bouret et al.\ (2003)
    for 3 SMC stars in NGC 346.

    The modified wind momentum - luminosity relation of all the SMC objects
    could be interpreted as showing a break-down at low luminosities or a different slope than the Galactic relation.
    The current sample of SMC stars may indeed indicate a steeper slope at least for giants and dwarfs, but the scatter is still too large to firmly establish this trend.
    However, the most recent hydrodynamical
    models (Vink et al.\ \cite{vink01}, Kudritzki \cite{kud02}, 
    Hoffmann et al.\ \cite{tadziu02}) do not predict such a change in the slope 
    between solar and SMC metallicities.
    Furthermore we present the first indications that some Galactic objects 
    have also low wind momenta comparable to the SMC dwarfs. This also tends to 
    exclude explanations based uniquely on metallicity.}

\item[$\diamond$]
    Possible explanations for a breakdown of the modified wind momentum 
    - luminosity relation at low luminosities are discussed.
    Ionic decoupling appears unlikely according to various estimates.
    A failure of the CAK parameterisation in high density atmospheres,    
    discussed by Owocki \& Puls (\cite{op99}), might be invoked to explain
    a lower acceleration in the transsonic region where the mass loss rate is
    set.

Although the physical mechanism leading to such weak winds remains
currently unknown, we speculate that the low mass loss rate
is probably intrinsically related to the youth of the stars, possibly
testifying of a phase of the ``onset'' of radiatively
driven winds in young O stars shortly after their formation.

\end{itemize}

Further studies of very young massive stars, Vz stars, and related objects 
with indications of weak winds will be of great interest to
attempt to understand these puzzling wind properties and
to provide interesting constraints on the development of stellar winds
in the early phases of massive star evolution or possibly even 
on the final phases of their birth.

\appendix
\section{Best fits}
\begin{figure}
\centerline{\psfig{file=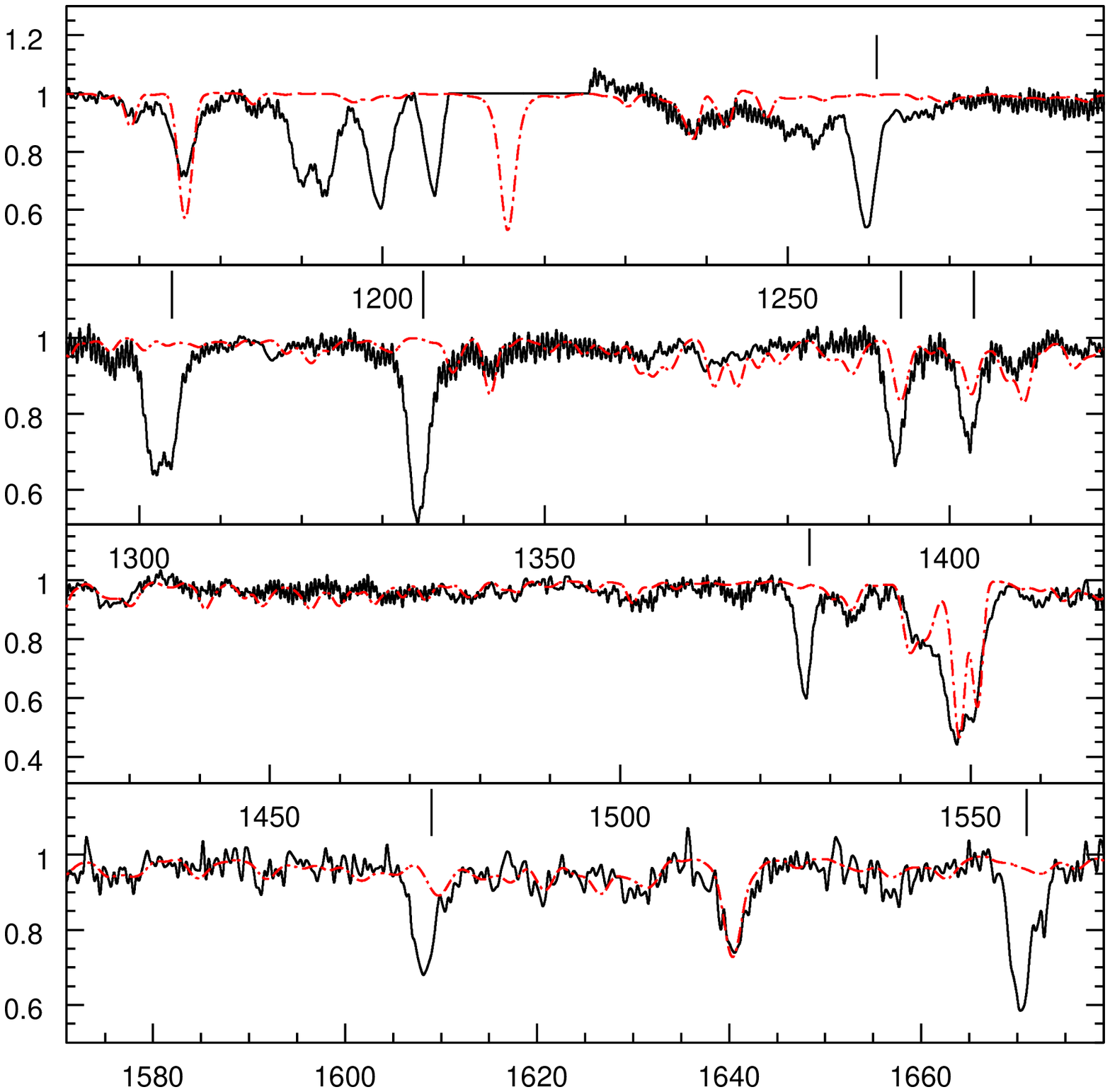,width=10cm,height=12cm}}
\caption{Best fit model for star 1. The model parameters are: \teff\
  = 38500 K, $\dot{M}$ = 10$^{-8.5}$ \msun yr$^{-1}$, v$_{\infty}$ =
  1500 km s$^{-1}$, \vsini\ = 200 km s$^{-1}$, f=0.01. The abundances
  are the
  following: n(He) = 0.1 n(H), C/C$_{\odot}$ = 1/10, N/N$_{\odot}$ =
  1/20, O/O$_{\odot}$ = 1/5, and Si, S and Fe abundances are 1/8 the
  solar values. Observation is the solid line and model is the
  dot-dashed line. The vertical lines indicate the main interstellar
  lines. 
}
\label{best_fit_s1}
\end{figure}

\begin{figure}
\centerline{\psfig{file=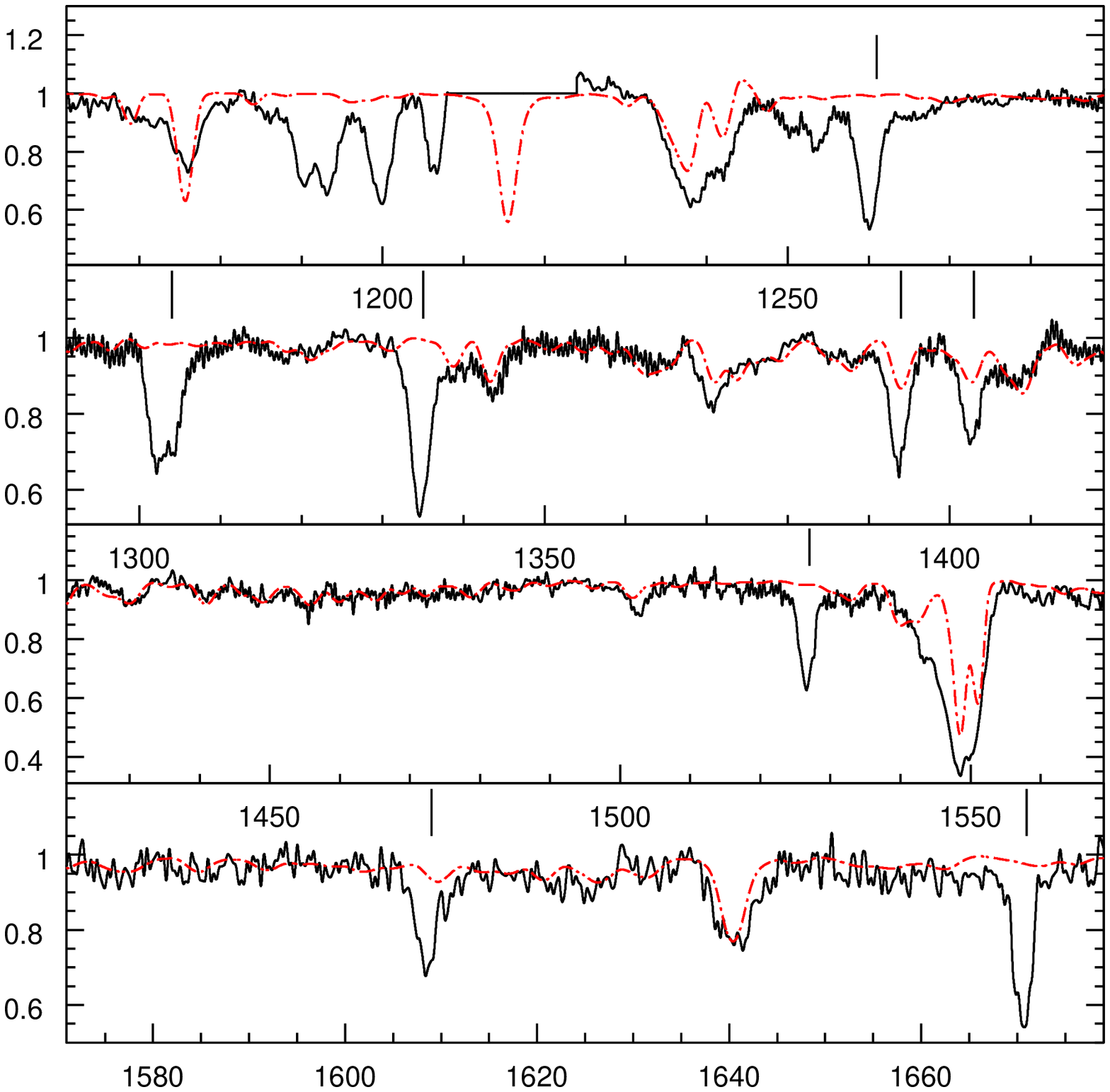,width=10cm,height=12cm}}
\caption{Best fit model for star 2. The model parameters are: \teff\
  = 40000 K, $\dot{M}$ = 10$^{-8.5}$ \msun yr$^{-1}$, v$_{\infty}$ =
  1800 km s$^{-1}$, \vsini\ = 300 km s$^{-1}$, f=0.01. The abundances are the
  following: n(He) = 0.1 n(H),and C, N, O, Si, S and Fe abundances are 1/8 the
  solar values. Observation is the solid line and model is the
  dot-dashed line. The vertical lines indicate the main interstellar
  lines. The poor fits of the \civ\ and \nv\ lines comes from the
  difficulty to have an important absorption without emission.
}
\label{best_fit_s2}
\end{figure}

\begin{figure}
\centerline{\psfig{file=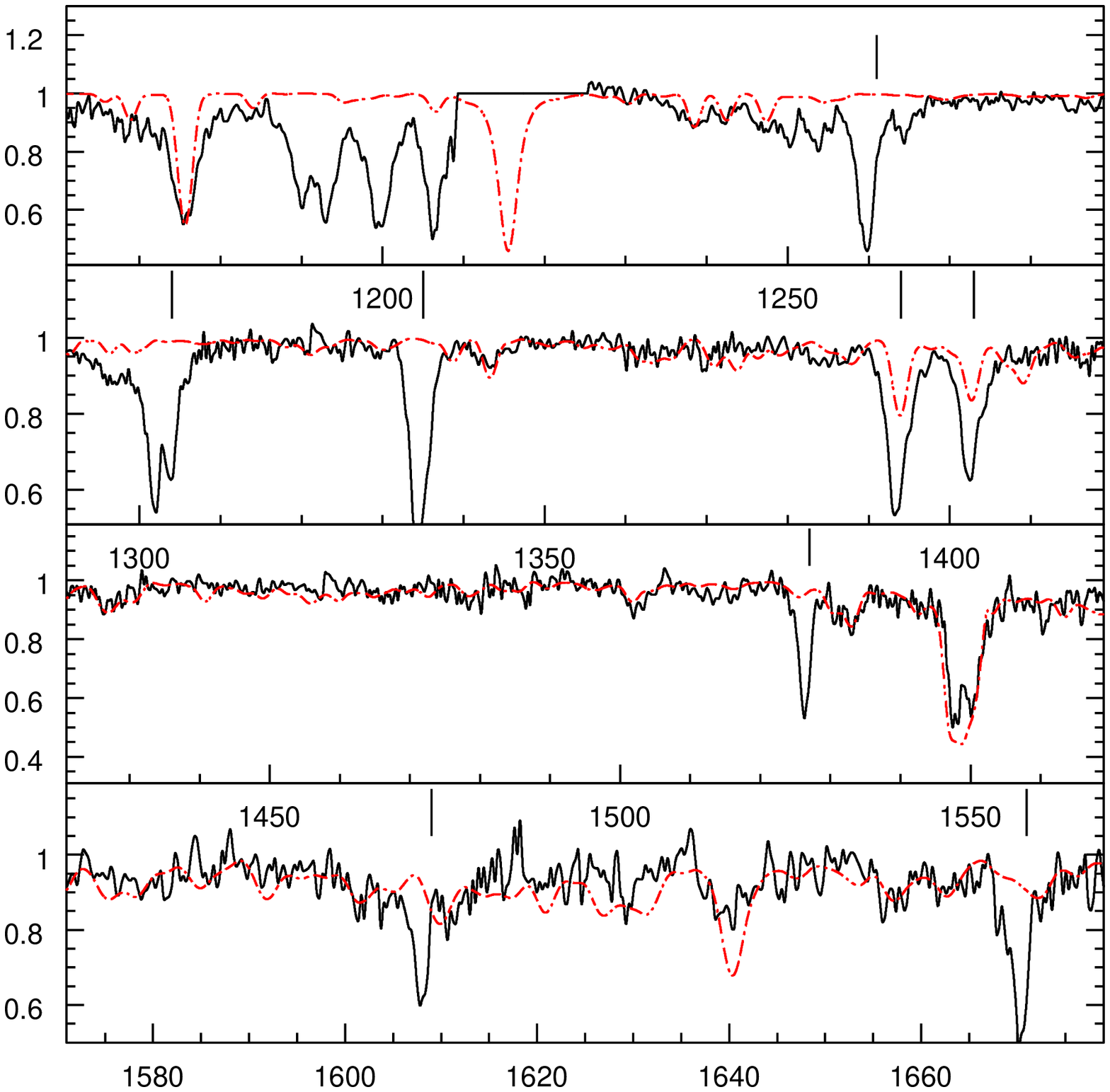,width=10cm,height=12cm}}
\caption{Best fit model for star 3. The model parameters are: \teff\
  = 36000 K, $\dot{M}$ = 10$^{-8.5}$ \msun yr$^{-1}$, v$_{\infty}$ =
  300 km s$^{-1}$, \vsini\ = 250 km s$^{-1}$. The abundances are the
  following: n(He) = 0.1 n(H), C/C$_{\odot}$ = 1/10, N/N$_{\odot}$ =
  1/20, O/O$_{\odot}$ = 1/5, and Si, S and Fe abundances are 1/8 the
  solar values. Observation is the solid line and model is the
  dot-dashed line. The vertical lines indicate the main interstellar
  lines.
}
\label{best_fit_s3}
\end{figure}

\begin{figure}
\centerline{\psfig{file=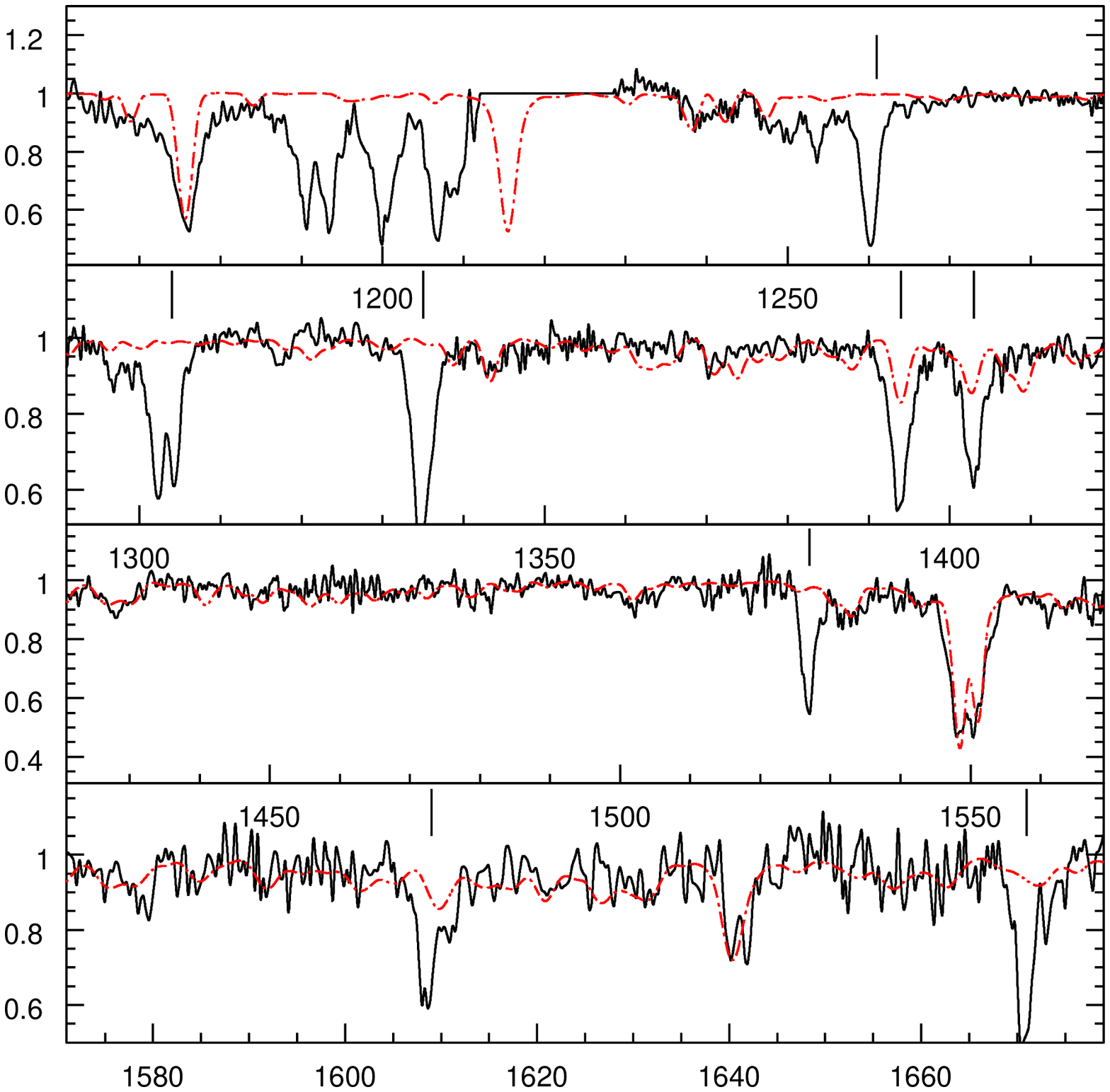,width=10cm,height=12cm}}
\caption{Best fit model for star 11. The model parameters are: \teff\
  = 37000 K, $\dot{M}$ = 10$^{-9.5}$ \msun yr$^{-1}$, v$_{\infty}$ =
  600 km s$^{-1}$, \vsini\ = 250 km s$^{-1}$. The abundances are the
  following: n(He) = 0.1 n(H), C/C$_{\odot}$ = 1/10, N/N$_{\odot}$ =
  1/20, O/O$_{\odot}$ = 1/5, and Si, S and Fe abundances are 1/8 the
  solar values. Observation is the solid line and model is the
  dot-dashed line. The vertical lines indicate the main interstellar
  lines.
}
\label{best_fit_s11}
\end{figure}

Fig.\ \ref{best_fit_s1}, \ref{best_fit_s2}, \ref{best_fit_s11} and
\ref{best_fit_s3} give the best fits achieved for star 1, 2, 11 and 3
respectively. For each figure, the model parameters are given and the
main interstellar lines are indicated. The interstellar CIV absorption 
has been added to the synthetic spectra as described in Sect.\
\ref{interstellar}. Even with this correction, the fit of the \civ\
line remain poor for star 2, showing the difficulty to produce a
significant absorption without emission. The normalisation of the
observed spectra below 1200 \AA\ is very uncertain so that any
comparison with models in this wavelength range is irrelevant.

\begin{acknowledgements}
We thank Jean-Claude Bouret, Luc Dessart, Claus Leitherer, Andr\'e Maeder, 
Georges Meynet and Stan Owocki
for useful discussions. We also thank Artemio Herrero and
Gerard Testor who kindly provided respectively the optical spectra of
10 Lac and the integrated optical spectra of SMC-N81 corrected for
nebular contamination. Artemio Herrero is also acknowledged for his
constructive comments as the referee of the paper.
The present results rely heavily on generous allocation of
computing time from the CALMIP and IDRIS centers. 
FM, DS, and MH-M thank the French ``Programme National de Physique Stellaire''
(PNPS) for support.
Part of this work was also supported by the French ``Centre National de Recherche
Scientifique'' (CNRS) and by the Swiss National Fund. 
\end{acknowledgements}

{}


\end{document}